\documentclass{aastex631}

\newcommand{\ts}{\textsuperscript}

\usepackage{url}
\usepackage{breakurl}
\def\UrlBreaks{\do\/\do-}

\newcommand{\AFFicrr}{\affiliation{Kamioka Observatory, Institute for Cosmic Ray Research, University of Tokyo, Kamioka, Gifu 506-1205, Japan}}
\newcommand{\AFFkashiwa}{\affiliation{Research Center for Cosmic Neutrinos, Institute for Cosmic Ray Research, University of Tokyo, Kashiwa, Chiba 277-8582, Japan}}
\newcommand{\AFFicrronly}{\affiliation{Institute for Cosmic Ray Research, University of Tokyo, Kashiwa, Chiba 277-8582, Japan}}
\newcommand{\AFFipmu}{\affiliation{Kavli Institute for the Physics and
Mathematics of the Universe (WPI), The University of Tokyo Institutes for Advanced Study,
University of Tokyo, Kashiwa, Chiba 277-8583, Japan }}
\newcommand{\AFFmad}{\affiliation{Department of Theoretical Physics, University Autonoma Madrid, 28049 Madrid, Spain}}

\newcommand{\AFFbu}{\affiliation{Department of Physics, Boston University, Boston, MA 02215, USA}}
\newcommand{\AFFuci}{\affiliation{Department of Physics and Astronomy, University of California, Irvine, Irvine, CA 92697-4575, USA }}
\newcommand{\AFFcsu}{\affiliation{Department of Physics, California State University, Dominguez Hills, Carson, CA 90747, USA}}
\newcommand{\AFFcnm}{\affiliation{Institute for Universe and Elementary Particles, Chonnam National University, Gwangju 61186, Korea}}
\newcommand{\AFFduke}{\affiliation{Department of Physics, Duke University, Durham NC 27708, USA}}
\newcommand{\AFFfukuoka}{\affiliation{Junior College, Fukuoka Institute of Technology, Fukuoka, Fukuoka 811-0295, Japan}}
\newcommand{\AFFgifu}{\affiliation{Department of Physics, Gifu University, Gifu, Gifu 501-1193, Japan}}
\newcommand{\AFFgist}{\affiliation{GIST College, Gwangju Institute of Science and Technology, Gwangju 500-712, Korea}}
\newcommand{\AFFuh}{\affiliation{Department of Physics and Astronomy, University of Hawaii, Honolulu, HI 96822, USA}}
\newcommand{\AFFicl}{\affiliation{Department of Physics, Imperial College London , London, SW7 2AZ, United Kingdom }}
\newcommand{\AFFkek}{\affiliation{High Energy Accelerator Research Organization (KEK), Tsukuba, Ibaraki 305-0801, Japan }}
\newcommand{\AFFkobe}{\affiliation{Department of Physics, Kobe University, Kobe, Hyogo 657-8501, Japan}}
\newcommand{\AFFkyoto}{\affiliation{Department of Physics, Kyoto University, Kyoto, Kyoto 606-8502, Japan}}
\newcommand{\AFFliv}{\affiliation{Department of Physics, University of Liverpool, Liverpool, L69 7ZE, United Kingdom}}
\newcommand{\AFFmiyagi}{\affiliation{Department of Physics, Miyagi University of Education, Sendai, Miyagi 980-0845, Japan}}
\newcommand{\AFFnagoya}{\affiliation{Institute for Space-Earth Environmental Research, Nagoya University, Nagoya, Aichi 464-8602, Japan}}
\newcommand{\AFFkmi}{\affiliation{Kobayashi-Maskawa Institute for the Origin of Particles and the Universe, Nagoya University, Nagoya, Aichi 464-8602, Japan}}
\newcommand{\AFFpol}{\affiliation{National Centre For Nuclear Research, 02-093 Warsaw, Poland}}
\newcommand{\AFFsuny}{\affiliation{Department of Physics and Astronomy, State University of New York at Stony Brook, NY 11794-3800, USA}}
\newcommand{\AFFokayama}{\affiliation{Department of Physics, Okayama University, Okayama, Okayama 700-8530, Japan }}

\newcommand{\AFFox}{\affiliation{Department of Physics, Oxford University, Oxford, OX1 3PU, United Kingdom}}

\newcommand{\AFFseoul}{\affiliation{Department of Physics, Seoul National University, Seoul 151-742, Korea}}
\newcommand{\AFFsheff}{\affiliation{Department of Physics and Astronomy, University of Sheffield, S3 7RH, Sheffield, United Kingdom}}
\newcommand{\AFFshizuokasc}{\affiliation{Department of Informatics in
Social Welfare, Shizuoka University of Welfare, Yaizu, Shizuoka, 425-8611, Japan}}
\newcommand{\AFFstfc}{\affiliation{STFC, Rutherford Appleton Laboratory, Harwell Oxford, and Daresbury Laboratory, Warrington, OX11 0QX, United Kingdom}}
\newcommand{\AFFskk}{\affiliation{Department of Physics, Sungkyunkwan University, Suwon 440-746, Korea}}

\newcommand{\AFFtodai}{\affiliation{Department of Physics, University of Tokyo, Bunkyo, Tokyo 113-0033, Japan }}
\newcommand{\AFFtit}{\affiliation{Department of Physics,Tokyo Institute of Technology, Meguro, Tokyo 152-8551, Japan }}
\newcommand{\AFFtus}{\affiliation{Department of Physics, Faculty of Science and Technology, Tokyo University of Science, Noda, Chiba 278-8510, Japan }}
\newcommand{\AFFtoronto}{\affiliation{Department of Physics, University of Toronto, ON, M5S 1A7, Canada }}
\newcommand{\AFFtriumf}{\affiliation{TRIUMF, 4004 Wesbrook Mall, Vancouver, BC, V6T2A3, Canada }}
\newcommand{\AFFtokai}{\affiliation{Department of Physics, Tokai University, Hiratsuka, Kanagawa 259-1292, Japan}}
\newcommand{\AFFtsinghua}{\affiliation{Department of Engineering Physics, Tsinghua University, Beijing, 100084, China}}
\newcommand{\AFFynu}{\affiliation{Department of Physics, Yokohama National University, Yokohama, Kanagawa, 240-8501, Japan}}
\newcommand{\AFFllr}{\affiliation{Ecole Polytechnique, IN2P3-CNRS, Laboratoire Leprince-Ringuet, F-91120 Palaiseau, France }}
\newcommand{\AFFbari}{\affiliation{ Dipartimento Interuniversitario di Fisica, INFN Sezione di Bari and Universit\`a e Politecnico di Bari, I-70125, Bari, Italy}}
\newcommand{\AFFnapoli}{\affiliation{Dipartimento di Fisica, INFN Sezione di Napoli and Universit\`a di Napoli, I-80126, Napoli, Italy}}
\newcommand{\AFFroma}{\affiliation{INFN Sezione di Roma and Universit\`a di Roma ``La Sapienza'', I-00185, Roma, Italy}}
\newcommand{\AFFpadova}{\affiliation{Dipartimento di Fisica, INFN Sezione di Padova and Universit\`a di Padova, I-35131, Padova, Italy}}
\newcommand{\AFFkeio}{\affiliation{Department of Physics, Keio University, Yokohama, Kanagawa, 223-8522, Japan}}
\newcommand{\AFFwinnipeg}{\affiliation{Department of Physics, University of Winnipeg, MB R3J 3L8, Canada }}
\newcommand{\AFFkcl}{\affiliation{Department of Physics, King's College London, London, WC2R 2LS, UK }}
\newcommand{\AFFwarwick}{\affiliation{Department of Physics, University of Warwick, Coventry, CV4 7AL, UK }}
\newcommand{\AFFral}{\affiliation{Rutherford Appleton Laboratory, Harwell, Oxford, OX11 0QX, UK }}
\newcommand{\AFFwu}{\affiliation{Faculty of Physics, University of Warsaw, Warsaw, 02-093, Poland }}
\newcommand{\AFFbcit}{\affiliation{Department of Physics, British Columbia Institute of Technology, Burnaby, BC, V5G 3H2, Canada }}
\newcommand{\AFFtohoku}{\affiliation{Department of Physics, Faculty of Science, Tohoku University, Sendai, Miyagi, 980-8578, Japan }}
\newcommand{\AFFicise}{\affiliation{Institute For Interdisciplinary Research in Science and Education, ICISE, Quy Nhon, 55121, Vietnam }}
\newcommand{\AFFilance}{\affiliation{International Laboratory for Astrophysics, Neutrino and Cosmology Experiment, Kashiwa, Chiba 277-8582, Japan}}
\newcommand{\AFFibs}{\affiliation{Institute for Basic Science, Yuseong-gu, Daejeon, Korea}}

\usepackage{xcolor}
\usepackage{soul}

\shorttitle{Pre-Supernova Alert System for Super-Kamiokande}

\begin{document}
\title{Pre-Supernova Alert System for Super-Kamiokande}

\correspondingauthor{L.~N.~Machado \\ \href{mailto:lucas.nascimentomachado@glasgow.ac.uk}{lucas.nascimentomachado@glasgow.ac.uk}}

\author{L.~N.~Machado}
\AFFnapoli

\author{K.~Abe}
\AFFicrr
\AFFipmu

\author{Y.~Hayato}
\AFFicrr
\AFFipmu
\author{K.~Hiraide}
\AFFicrr
\AFFipmu
\author{K.~Ieki}
\AFFicrr
\author{M.~Ikeda}
\AFFicrr
\AFFipmu
\author{J.~Kameda}
\AFFicrr
\AFFipmu
\author{Y.~Kanemura}
\AFFicrr

\author{R..~Kaneshima}
\AFFicrr
\author{Y.~Kashiwagi}
\AFFicrr

% ...

\collaboration{225}{(The Super-Kamiokande Collaboration)}


\begin{abstract}
In 2020, the Super-Kamiokande (SK) experiment moved to a new stage (SK-Gd) in which gadolinium (Gd) sulfate octahydrate was added to the water in the detector, enhancing the efficiency to detect thermal neutrons and consequently improving the sensitivity to low energy electron anti-neutrinos from inverse beta decay (IBD) interactions. SK-Gd has the potential to provide early alerts of incipient core-collapse supernovae through detection of electron anti-neutrinos from thermal and nuclear processes responsible for the cooling of massive stars before the gravitational collapse of their cores. These  pre-supernova neutrinos emitted during the silicon  burning phase can exceed the energy threshold for IBD reactions. We present the sensitivity of SK-Gd to pre-supernova stars and the techniques used for the development of a pre-supernova alarm based on the detection of these neutrinos in SK, as well as prospects for future SK-Gd phases with higher concentrations of Gd. For the current SK-Gd phase, high-confidence alerts for Betelgeuse 
could be issued up to nine hours in advance of the core-collapse itself.
\end{abstract}


\section{Introduction} \label{sec:intro}
The first observation of neutrinos produced outside of our solar system happened on February 23\ts{rd}, 1987, due to a neutrino burst from a supernova explosion known as SN1987A in the Large Magellanic Cloud, located approximately 50 kiloparsecs (163,000 light years) from Earth. While this supernova -- the first to be observed in the vicinity of our own galaxy since the invention of the telescope -- was initially detected via its emitted visible light, signals from the corresponding neutrino burst were found in recorded data to have arrived several hours earlier~\citep{1987A, 1987B, 1987C, 1987D}; the neutrinos are generated before the majority of the electromagnetic radiation. SN1987A remains the only observation of supernova neutrinos in history. Since then, a variety of more modern experiments with such advantages as greater longevity, increased target mass, lower energy threshold, and improved energy and timing resolution have come online (e.g. Super-Kamiokande~\citep{FUKUDA2003418},  IceCube~\citep{Aartsen_2017}, KamLAND~\citep{Eguchi_2003}, SNO+~\citep{Albanese_2021}, etc.), each hoping to efficiently detect supernova neutrino bursts and possibly provide early alerts for the delayed electromagnetic signal.  

The Super-Kamiokande (SK) experiment~\citep{FUKUDA2003418}, located in the Kamioka mine in Japan, has the potential of detecting low energy (above a few MeV) neutrino bursts from supernova explosions in our galaxy and its surroundings. The neutrino elastic scattering interaction is a subdominant (around 3\% of all events) but important interaction channel for low energy neutrinos in SK, allowing the detector supernova pointing capability by reconstructing the direction of scattered electrons. SK is currently the largest and most sensitive SN neutrino detector in the world and it has an active real-time supernova neutrino burst monitor~\citep{2016snwatch}. In 2020, the water in the SK detector was loaded with gadolinium (Gd) to increase the project's sensitivity to low energy electron anti-neutrinos ($\bar{\nu}_e$) by improving the identification of neutrons resulting from their most likely (about 85\%) interaction in water: inverse beta decay (IBD).

But not only supernova bursts can provide alerts for supernova explosions. Prior to the collapse of their cores, very massive stars nearing the end of their lives are supported by the nuclear fusion of heavy nuclei.  These are commonly known as pre-supernova stars; their main cooling mechanism is via neutrino emission.  Pairs of neutrinos and anti-neutrinos of all flavors are produced during this phase~\citep{odrzywolek2004neutrinos}, and the enhanced sensitivity to low energy $\bar{\nu}_e$ in SK-Gd allows the detection of neutrinos emitted during the terminal silicon (Si) burning period.

A potential world-first detection of pre-supernova neutrinos by SK-Gd would help to resolve open questions regarding stellar evolution models as well as provide valuable input regarding the longstanding issue of the neutrino mass hierarchy~\citep{kato_annual_review}. These new detection capabilities have also motivated the creation of a pre-supernova alert system for the experiment. SK-Gd's sensitivity to pre-supernova neutrinos and details of its new  alert system are presented in this article.

\section{The Super-Kamiokande Experiment}\label{sec:sk_experiment}
The SK experiment is a 50 kiloton (kt) water Cherenkov detector located in the Kamioka mine in Japan~\citep{Super-Kamiokande:2002weg}. It was designed to study neutrinos with energies from a few MeV to few hundreds of GeV, produced by a variety of both natural and artificial sources. The detector consists of a cylindrical stainless steel tank with a 39.3 meter diameter and a 41.4 m height, divided into an inner detector (ID) and an outer detector (OD). The ID is responsible for the event detection, with over 11,000 20-inch photomultiplier tubes (PMTs) and it has a volume of 32~kt, although the usual fiducial volume (FV) used in SK analyses is 22~5 kt. The OD has a thickness of about 2~m and it is composed of 1,885 8-inch PMTs, facing the outside of the detector to reduce entering cosmic ray-induced backgrounds.

From July 14\ts{th}, 2020, SK officially started the SK-Gd phase, in which 13 tons of gadolinium sulfate octahydrate $\rm Gd_2(\rm SO_4)_3\cdot \rm 8H_2O$\ was added to the water in the detector, achieving a concentration of 0.01\% Gd by mass~\citep{ABE2022166248}. Gadolinium sulfate octahydrate is an easily soluble compound and it has good transparency to the Cherenkov light. The SK-Gd project was first proposed as GADZOOKS! by~\citep{PhysRevLett.93.171101} to improve the identification of neutrons resulting from processes including inverse beta decay (IBD), the main interaction channel for low energy $\bar{\nu}_e$. IBD interactions have in their final state a prompt positron, which is directly detected through Cherenkov radiation, and a delayed neutron, which travels a short distance before being thermalized in water and then captured by nuclei. These nuclei de-excite by emitting $\gamma$-rays that are detected mainly due to Compton-scattered electrons producing Cherenkov light. Among all the natural occurring elements, Gd has the largest thermal neutron cross section capture, of approximately 49,000 barns.

The de-excitation $\gamma$-rays have different energies depending on the capture nuclei. A neutron capture on hydrogen emits a single 2.2 MeV $\gamma$-ray, which has low detection efficiency since its Compton electrons are close to or below the Cherenkov threshold; in pure water the neutron capture time constant is approximately $200$ ${\mu}s$. With small amounts ($\geq$ 0.01\% by mass) of gadolinium  dissolved in the water of the detector, the majority of thermal neutron captures will be on the Gd since it has a neutron capture cross section about 100,000 times larger than that of hydrogen. These captures on Gd emit easily detectable $\gamma$-ray cascades of about 8 MeV~\citep{PhysRevLett.93.171101}. The current phase of SK-Gd corresponds to a loading of 0.01\% Gd by mass and, as a consequence, approximately 50\% of the neutron captures are on Gd. The presence of Gd  shortens the neutron capture time constant, which for the present loading has been measured to be $115 \pm 1$ ${\mu}s$~\citep{ABE2022166248}. In the next loading phase -- planned to begin in mid-2022 -- the Gd concentration will be increased to 0.03\% to yield $\sim$75\% captures on Gd. The final goal is to have SK loaded with 0.1\% Gd, which will result in approximately 90\% of the captures being on Gd.

The search for IBD events is characterized by the coincidence between the prompt positrons and delayed neutrons. This signature is used to distinguish background events, since it is unlikely that uncorrelated events produce the same pattern. The efficient neutron identification provided by gadolinium brings benefits to different analyses in SK such as supernova neutrinos, proton decay and atmospheric neutrino studies, and in the search for signals  from nuclear power reactors and the Diffuse Supernova Neutrino Background (DSNB), which is the as-yet unobserved~\citep{PhysRevLett.93.171101, BEACOM_DSNB} integrated neutrino flux from all core-collapse supernova (CCSN) throughout the history of the universe. Another possible first-time detection of SK-Gd would be the neutrinos coming from the silicon (Si) burning phase of stars preceding the CCSN, also known as pre-supernova neutrinos.

The EGADS (Evaluating Gadolinium’s Action on Detector Systems) detector has been used to simulate the Gd loading in SK~\citep{MARTI2020163549}. Also located in the Kamioka mine, EGADS was designed to be a smaller prototype of SK-Gd; it is a stainless steel tank with 200 tons of water and 240 PMTs installed on the walls, and uses the same electronics as SK. Among other important results, prior to the loading of gadolinium into SK's water EGADS demonstrated the ability to maintain ultrapure water-style  transparency in the presence of Gd, a viable method of safely removing the Gd from the detector when desired, and the additions to the existing SK water system needed in order to prepare for Gd loading. The EGADS detector is currently running with a concentration of 0.03\% Gd by mass in order to validate  operating conditions and techniques for the next planned SK-Gd phase. 

\section{Pre-Supernova Neutrinos}
Massive stars, which are stars with initial masses greater than 8 solar masses ($M_\odot$) at the Zero Age Main Sequence (ZAMS), often\footnote{Alternatively, massive stars may collapse directly to a black hole with no supernova explosion~{\citep{O_Connor_2011}.}} end their lives in a CCSN. Prior to their cores collapsing -- as massive stars burn most of their hydrogen into helium -- the increasing density and temperature of their cores allow the fusion of heavier nuclei~\citep{Woosley1978TheEA}. With the ignition of carbon burning, massive stars are classified as neutrino-cooled stars~\citep{Arnettbook}; in this phase the amount of neutrino emission from thermal and nuclear processes increases significantly, becoming the main cooling mechanism of these stars. A neutrino-cooled star lasts hundreds of years, reaching a neutrino luminosity of about $10^{12}$ solar luminosity ($L_\odot$), while the photon luminosity is only $10^{5}L_\odot$ at this stage~\citep{odrzywolek2004neutrinos}.

When the neutrino-cooled stage begins, massive stars typically\footnote{Some light massive stars can proceed with oxygen-neon-magnesium burning \citep{Giunti:2007ry}.} proceed with nuclear fusion of helium (He), carbon (C), oxygen (O), neon (Ne) and silicon (Si). The burning of these elements starts from the core and then propagates toward the edges of the star in concentric shells~\citep{Woosley1978TheEA} due to differing  densities and temperatures within the stellar volume. Closer to the onset of the gravitational collapse, the final structure of the neutrino-cooled star is an iron core surrounded by shells containing the products of the sequence of elemental nuclear burning phases. At this point, neutrino-cooled stars are commonly called pre-supernova stars.

Although many processes contribute to the neutrino emission from pre-supernova stars, at the very high temperatures at this stage  the electron-positron annihilation process generating thermal neutrinos (Eq. \ref{pair_annihilation}) is the star's dominant form of cooling: 
\begin{equation}
    e^+  + e^- \rightarrow  \nu_x + \bar{\nu}_x,
    \label{pair_annihilation}
\end{equation}
where $x=e, \mu, \tau$.

\begin{figure}[htb!]
    \centering
    \includegraphics[scale=0.6]{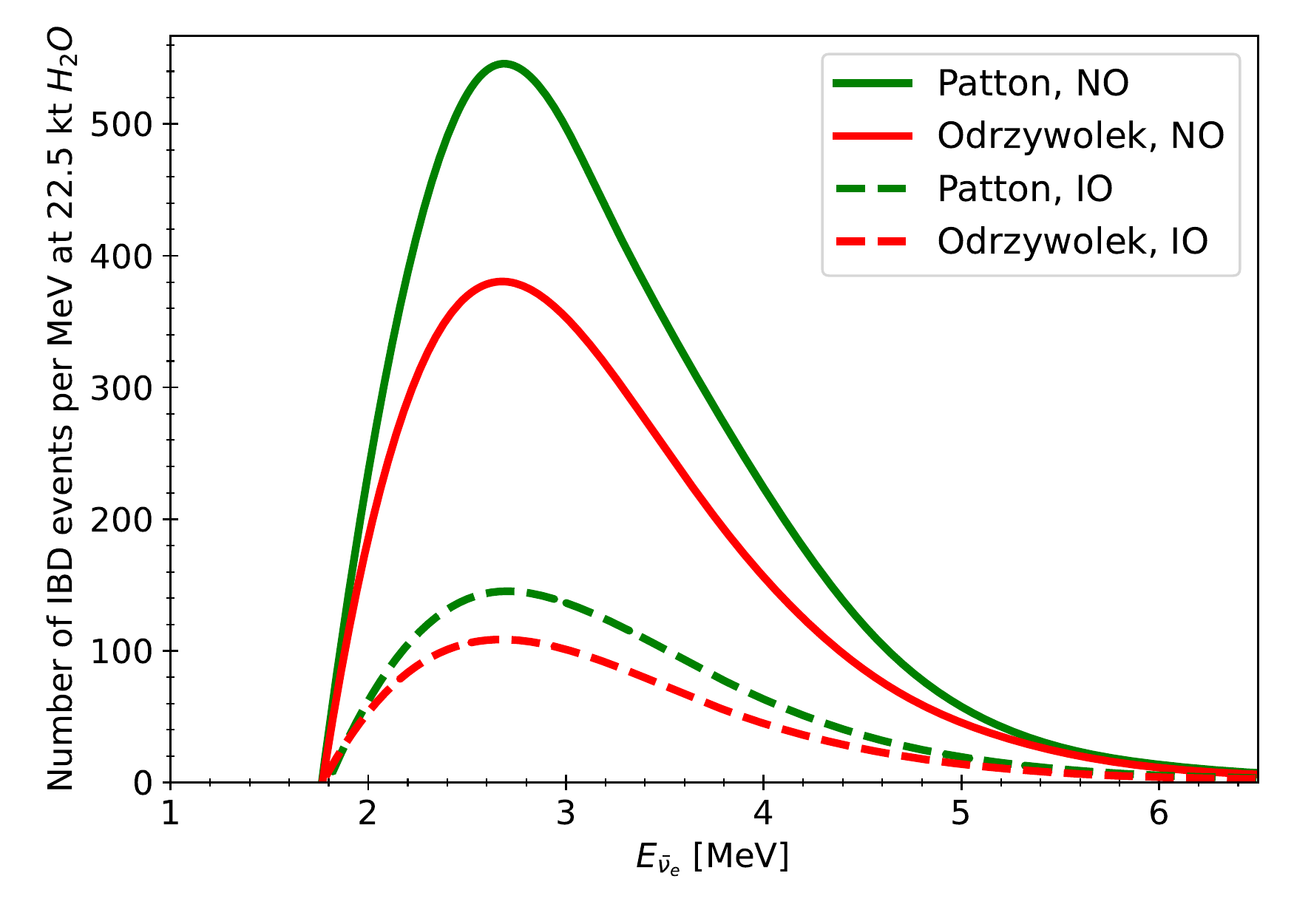}
    \caption{Number of pre-supernova IBD interactions in the 22.5 kt SK FV integrated over the last 10 hours prior to the CCSN as a function of the $E_{\bar{\nu}_e}$. The Betelgeuse-like models consider stars with initial masses of 15 M$_{\odot}$ located 150 parsecs away from Earth, for both normal neutrino mass ordering (NO) and inverted neutrino mass ordering (IO).}
    \label{fig:number_IBD_nu_energy}
\end{figure}

The Si-burning phase, which is expected to last for just a few days at temperatures of approximately $3 \times 10^9$ K~\citep{Woosley1978TheEA, odrzywolek2004neutrinos}, creates an iron core whose inability to generate energy via fusion into still heavier elements can then initiate the CCSN. The electron anti-neutrinos emitted at the Si-burning stage, which correspond the 1/3 of the anti-neutrino flux~\citep{odrzywolek2007}, can exceed the energy threshold of IBD reactions, making their detection possible in SK-Gd. Figure~\ref{fig:number_IBD_nu_energy} shows the expected number of IBD interactions in SK for different Betelgeuse-like pre-supernova models during the final 10 hours prior to core collapse for 15 M$_{\odot}$ stars located 150 parsecs away from Earth.

Pre-supernova $\bar{\nu}_e$s have never been observed before;  their detection would not only provide unique, unperturbed  information detailing otherwise hidden stellar interiors, but also contribute evidence regarding which neutrino mass hierarchy is the correct one~\citep{2019gun, kato_annual_review}. Being able to detect them also opens the possibility of creating a pre-supernova alarm for SK, potentially delivering alerts hours before the arrival of any other CCSN signals since the emission of pre-supernova neutrinos takes place over a very long timescale compared to the subsequent supernova burst neutrinos upon which most SN alarms are based.

\subsection{Pre-Supernova Models}
In addition to pair annihilation $ e^+ e^-  \longrightarrow  \bar{\nu} \nu$ for the cooling of massive stars, beta processes are expected to also contribute significantly at the pre-supernova stage, with an average more energetic $\bar{\nu}_e$ flux~\citep{2017patton}. Due to the large interaction cross section for IBD at both the typical pair annihilation and beta process energies, only electron anti-neutrinos from the silicon burning phase are expected to be detected in SK~\citep{odrzywolek2004neutrinos}.

Following previous analyses~\citep{2019charles}, in order to perform estimations of the expected signal from pre-supernova neutrinos in SK, two models for the thermodynamics of stellar evolution were used: \citep{Odrzywolek:2010zz} and \citep{2017patton}. Both models provide online data sets for the calculation of anti-neutrino emission during the pre-supernova stage. The first, earlier model from~\citep{Odrzywolek:2010zz} assumes that the entire neutrino flux comes from pair annihilation. For the nuclear isotopic composition of the star, this model assumes a nuclear statistical equilibrium (NSE), which is a treatment only dependent on the temperature, density and electron fraction, making a flux estimate by post-processing an already existing stellar model. The second model~\citep{2017patton} includes a more complete evaluation of the neutrino flux from the pre-supernova stars, considering the contribution of individual isotopes to the stellar evolution, which affects the neutrino emission rate from weak nuclear processes. The model also includes contributions not only from pair annihilation, but also from other thermal and nuclear processes such as plasmon decay, photoneutrino process, $\beta$-decay, and electron capture.

For signal estimations, the flavor oscillations of electron anti-neutrinos as they travel through the stellar volumes were taken into account. The description of oscillations in matter is called the Mikheyev-Smirnov-Wolfenstein (MSW) effect~\citep{smirnov2003msw}. The ratio by which the flux of electron anti-neutrinos is changed depends on the electron number density of the star and the neutrino mass ordering (as yet unknown), which changes the mixing parameters. Different transition probabilities are assumed for normal and inverted neutrino mass ordering to account for this change in the ratio of electron flavor neutrinos due to the dense stellar medium in addition to the effects of oscillations in vacuum.

\section{Analysis Strategy}
The detection of low energy electron anti-neutrinos from pre-supernova stars in SK is mainly through the IBD interaction, where the products are a prompt positron and a delayed neutron. Searches for IBD events consider the coincidence of the products of the interaction.

Previous analysis~\citep{2019charles} of the sensitivity of SK to pre-supernova stars considered the detection of two different detection channels; in addition to the search for prompt and delayed parts, the search for just the delayed neutrons individually was also used to calculate the expected sensitivity. For this analysis, however, only coincidence events are considered since, due to the higher rate of single neutron events, the further development of an online alert system would be negatively affected by a significant increase of the processing time. For this reason, techniques were developed to optimize the online selection of only coincidence events, while the single neutron events will be utilized in a planned future improvement of the alert system to post-check data in case of potential alerts.

\subsection{Event Simulation}\label{sec:event_simulation}
In order to simulate the IBD events, and following what was done in~\citep{2019charles}, two different $\gamma$-ray emission models from neutron capture on Gd are used: the generic liquid scintillator simulator GLG4sim~\citep{chooz_ref}, and a second model created by a group at Okayama University using data from the Accurate Neutron-Nucleus Reaction Measurement Instrument (ANNRI) spectrometer at the Japan Proton Accelerator Research Complex (J-PARC). The ANNRI model was based on measurements of $\gamma$-ray properties with an array of germanium detectors at the facility~\citep{anri_ref_sim}. The GLG4sim model uses GEANT4~\citep{AGOSTINELLI2003250} support for neutron capture on Gd, but also includes spectral information regarding high energy $\gamma$-rays, which the original GEANT4 does not consider~\citep{chooz_ref}. The ANNRI model uses GEANT4 Monte Carlo simulation and the spectrometer data to describe the $\gamma$-ray spectrum from neutron captures on Gd~\citep{10.1093/ptep/ptz002}. The models assume an isotropic distribution for the angles of $\gamma$-ray emission. Both are included in this analysis to provide systematic uncertainties for the pre-supernova neutrino sensitivity results in SK. 

\begin{figure}[htb!]
    \centering
    \includegraphics[scale=0.65]{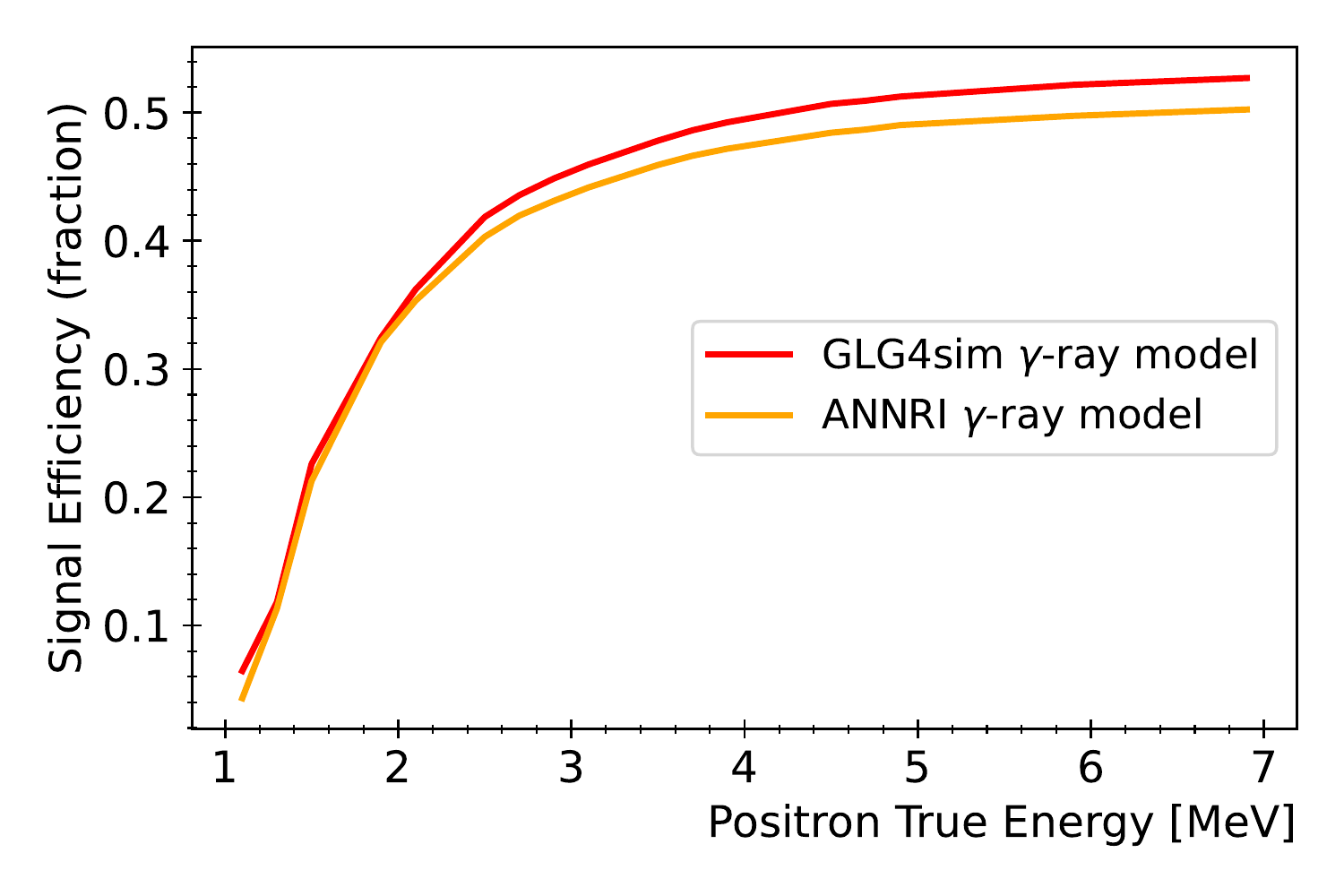}
    \caption{Selection efficiency versus positron true energy for SK with 0.01\% Gd for the two $\gamma$-ray models used for analysis GLG4sim~\citep{chooz_ref} and ANNRI~\citep{anri_ref_sim}.}
    \label{fig:selection_eff}
\end{figure}

To reproduce IBD coincidence events, simulation of prompt positrons and delayed neutrons were performed and put in coincidence in time and distance to each other. The SK Monte Carlo software was employed to simulate positron events and $\gamma$-ray events from the neutron captured on Gd, using the GLG4sim and ANNRI models. These coincidence events were simulated in different positions of the inner detector, with positron energies ranging between 0.8~MeV and 7.0~MeV. Figure~\ref{fig:selection_eff} shows the signal efficiency as a function of positron true energy for the simulations using both $\gamma$-ray models for the event selection that will be described in Sections~\ref{sec:WITtrigger} and \ref{sec:event_selection}.

\subsection{Backgrounds}
The sensitivity to pre-supernova neutrinos at SK with 0.01\% Gd was evaluated using Monte Carlo and SK data prior to the first Gd loading to calculate the contributions from different background sources, which later were compared to SK-Gd data. The main backgrounds are:

\begin{itemize}
    \item Reactor and geo electron anti-neutrinos in the same energy region as pre-supernova neutrinos.  These are true, irreducible physics backgrounds.
    \item Trace amounts of radioactive isotopes mixed in along with the Gd. After being distributed in the detector these radioactive impurities could contribute to backgrounds from $(\alpha,n)$ and spontaneous fission processes.    
    \item Accidental coincidences, which are intrinsic detector backgrounds due to radioactive decays from the detector materials, PMT dark noise, and uncorrelated events that  randomly occur close in time and distance. These backgrounds were evaluated by processing around 7000 hours of data from different SK data periods.
    \item Spallation, due to cosmic rays muons. Most of these low energy decays of unstable nuclei can be eliminated as they are correlated in time and space with energetic muons.
\end{itemize}

The main background sources for pre-supernova neutrinos are the reactor neutrinos, primarily a result of the activity of Japanese nuclear power reactors, and geoneutrinos, which come from the decay of natural radioisotopes inside the Earth. These are  irreducible backgrounds, in the same energy range as pre-supernova neutrinos. The web page \url{geoneutrinos.org} has an application to evaluate reactor and geoneutrino fluxes, giving the rate and energy spectrum of anti-neutrino interactions at any location on the planet~\citep{dye2021global}.  The application uses the IAEA's database Power Reactor Information System (PRIS)~\citep{international1989iaea} to account for each reactor's activity during specific periods. To give the best possible estimations of the current reactor and geoneutrino backgrounds, when applying this tool to the pre-supernova alarm the most up-to-date information from \url{geoneutrinos.org} is used and further corrected with the most recently updated information regarding reactor activity in Japan.

Backgrounds due to radioactive contamination of the Gd loading come mainly from ${}^{18}$O($\alpha$,n)${}^{21}$Ne$^*$ and ${}^{17}$O($\alpha$,n)${}^{20}$Ne$^*$ processes, mainly a consequence of contamination with ${}^{235}$U series isotopes that are $\alpha$ emitters. The decay of neon isotopes results in pairs of neutrons and the signal of one of the neutrons can be mistaken for a positron, thereby becoming a background for coincidence IBD events. Since the spontaneous fission of ${}^{238}$U also produces one or more neutrons per fission, this also can contribute to the overall background level.

Muons from cosmic-rays create unstable daughter nuclei through spallation, which can in turn produce $\beta$-delayed neutrons (neutrons associated with the beta decay of the fission products). In particular, $\beta$-ray energy for the decay of nitrogen-17 is in the energy range of interest for this analysis. 

Initially, all of the described backgrounds were evaluated with Monte Carlo, apart from the accidental coincidences background which was calculated from data. When SK-Gd data became available, these backgrounds were more rigorously quantified after processing 2000 hours of data. After the event selection, described in Section~\ref{sec:event_selection}, the remaining background rate is taken into account to calculate the sensitivity to pre-supernova neutrinos in SK. 

\subsection{Trigger Considerations}\label{sec:WITtrigger}
The pre-supernova alert system is integrated into an independent trigger in SK called the Wide-band Intelligent Trigger (WIT)~\citep{CARMINATI2015666,ElnimrWIT}. WIT is a system designed to extend the sensitivity of SK to lower energy events using parallel computing to reconstruct vertices in real time, discarding events that are not well reconstructed or very close to the walls of the detector. The WIT system consists of more than 400 hyper-threaded cores spread over ten online computers that work in parallel with each other. The system will be significantly upgraded during 2022 with a batch of new computers containing hundreds of additional cores.

The WIT system receives raw data blocks containing about 23~milliseconds of data each from data acquisition machines; these
blocks are then distributed to the ten online computers to search for signals (11~PMT hits within 230~nanoseconds) above expected dark noise levels (12~hits) inside these blocks.

The hits passing this procedure proceed to another filter called STORE (Software Triggered Online Reconstruction of Events). STORE looks for PMT hits originating from a single vertex by considering the Cherenkov emission of low energy electrons a point source, a reasonable assumption since these electrons can only travel a few centimeters in water~\citep{CARMINATI2015666}, which is much smaller than SK's vertex  resolution.  STORE selects the largest set of hits that are consistent with being from a point source. Another trigger condition applied requires that for any two PMT hits the time difference between them must be less than or equal to the time required to travel between them at the speed of light in water. From these selected hits, STORE gathers four-hit combinations, using their arrival times to calculate vertex position, in which the time residuals ((time of the PMT hit) - (time of flight) - (time of the test vertex)) of the four hits must be zero.

The selected hits from STORE go through a fast vertex reconstruction called ClusFit, which eliminates isolated hits to reduce effects of reflected and scattered light and dark noise~\citep{CARMINATI2015666}. The fast online vertex reconstruction in WIT is concluded by applying BONSAI (Branch Optimization Navigating Successive Annealing Iterations)~\citep{SmyBONSAI} to the selected events from Clusfit that are at least 1.5~m away from any PMT.

BONSAI uses events within 500~ns before and 1000~ns after the trigger time, performing a maximum likelihood fit to all arrival times of PMT hits inside this window, testing for each one a vertex hypothesis and choosing the ones with maximized likelihood as the reconstructed vertices. BONSAI saves the 1500~ns events that are inside the fiducial volume (2~m away from the inner PMT wall) and that have at least 10 hits with time residuals within $-6$~ns and $+12$~ns~\citep{spallation_skIV}.

The reconstructed events are sent to WIT's organizer machine, which is a computer responsible for sorting the events in time, gathering about 90~seconds worth of data into one file. The sorted events are then transferred to the offline computers for further processing by the various offline SK physics analyses.

\subsection{Event Selection}\label{sec:event_selection}
To select the IBD pre-supernova events, multivariate methods are used, in particular Boosted Decision Tree (BDT)~\citep{articleBDT}, which is applied at two different levels of the data processing for reasons of both speed and efficiency. While one of the BDTs is used as the final selection of pre-supernova events, the other BDT, called BDT$_{online}$, is applied to search for IBD candidate events in the WIT data.

The initial search for IBD pair candidates is performed using the number of hit PMTs and the reconstructed position and time of events. This information is available online when data is being saved by the WIT trigger, looping over all events passing the fast online reconstruction. For each prompt event candidate, a delayed candidate event is searched in a time window from -17~$\mu s$ to 290~$\mu s$ relative to the prompt candidate (each hardware trigger is 17~$\mu s$ long). If the delayed candidate has a minimum number of hit PMTs, the time between the prompt and delayed candidates, dT, as well as the distance between their reconstructed positions, dR, are calculated.

As described in Section~\ref{sec:sk_experiment}, IBD events are very well characterized by the spatial and timing coincidences between the prompt and delayed signals, represented by the  variables dR (distance) and dT (time). When comparing accidental coincidences with signal, these variables have very different distributions. However, a selection using only this information is not enough to sufficiently reduce backgrounds. For that reason BDTs were implemented in the analysis.

\begin{figure}[htb!]
\gridline{\fig{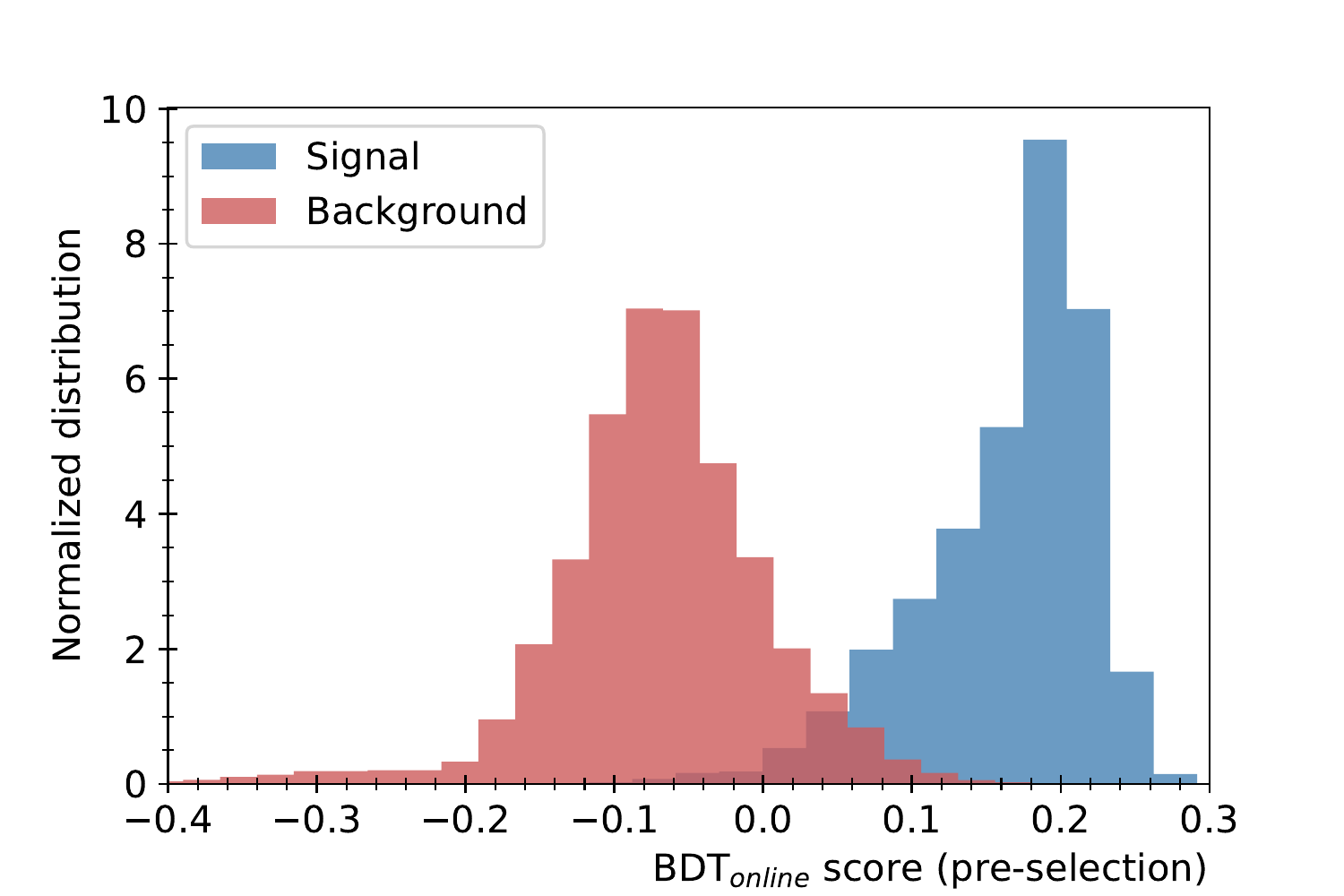}{0.45\textwidth}{(a)}
          \fig{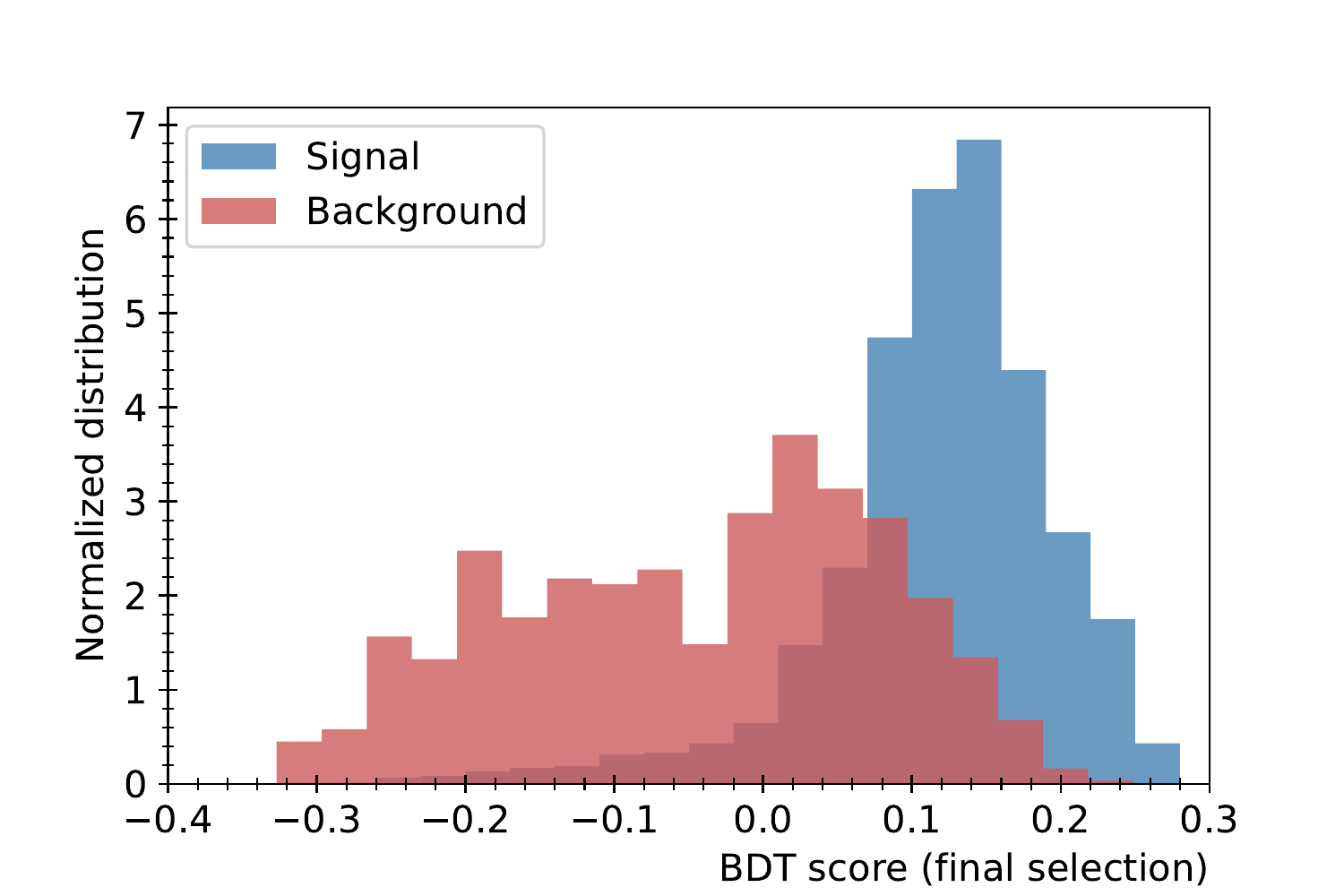}{0.45\textwidth}{(b)}
          }
\caption{Signal-background separation for the two Boosted Decision Tree classifiers used for (a) pre-selection (BDT$_{online}$) and (b) final selection.}
\label{fig:bdts}
\end{figure}

BDT$_{online}$, applied during the initial search for signal events, was trained based only on the variables available online from fast WIT online vertex reconstruction; BDT$_{online}$ is used at this early realtime processing stage as a pre-selection of potential IBD candidates. The final selection of events is done by applying cuts in dR, dT, and to a score resulting from another BDT classifier, which is based on angular distribution of hits, reconstructed energy and quality, and distance from events to the detector wall. Figure~\ref{fig:bdts} shows the signal and background separation for both BDTs. Both BDTs are trained using random subsets of SK-Gd data as background and a portion (about 30\%) of the generated IBD coincidence events using the GLG4sim model, as described in Section \ref{sec:event_simulation}, as signal.

Using two BDTs in the selection process is justified by the need for having fast processing time in the alert system, since it is an online system that constantly and uninterruptedly processes many data sets, running statistical evaluations to make alarm decisions and send alerts. For this reason, the system needs to process data as quickly as possible, but without losing detection efficiency. BDT$_{online}$ improved the speed of the alert system by filtering pair candidates more efficiently, reducing the number of background events being carried through the full data reduction. The combination of the two BDTs ultimately resulted in better sensitivity to pre-supernova neutrinos than using only one of them.

Analyzing 2000 hours of SK-Gd data, optimizations to the selection were performed and the backgrounds' contributions were re-evaluated in order to quantify the contributions from the different sources and better predict the sensitivities of the current and future SK-Gd loading phases.  These results are described in the next section.
\section{Sensitivity to Pre-supernova Neutrinos}\label{sec:sensitivity}
Previous sensitivity results for detection of pre-supernova neutrinos in SK with 0.1\% Gd were given in~\citep{2019charles}. However, the first -- and current, as of this writing -- phase of SK-Gd has a concentration of 0.01\% Gd, and the development of the pre-supernova alert system required a dedicated response model for the current phase. In this section, the estimated  sensitivity to pre-supernova neutrinos is presented. General details about the alert system, its expected early warning times, and its coverage of our galaxy are covered in Section~\ref{sec:alert_system}.

\begin{figure}
    \centering
    \includegraphics[scale=0.2]{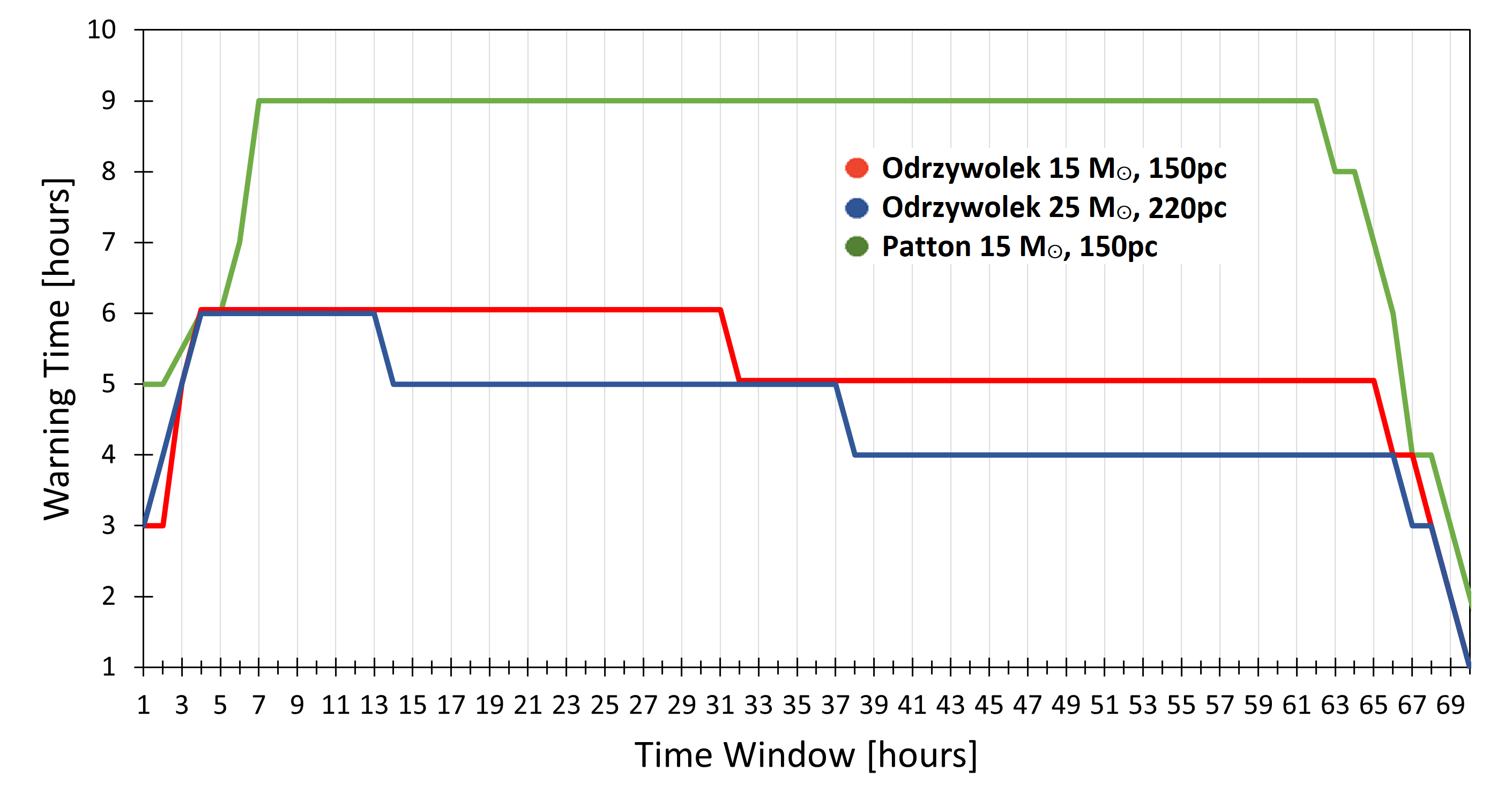}
    \caption{Expected warning times for Betelgeuse-like models  evaluated as a  function of the chosen time window to perform statistical evaluations in hourly steps. The estimations consider the neutrino flux from the last 70 hours before  core-collapse for stars with 15~M\textsubscript{\(\odot\)} at 150~parsecs and 25~M\textsubscript{\(\odot\)} at 220~parsecs following the model \citep{Odrzywolek:2010zz}, and 15~M\textsubscript{\(\odot\)} stars at 150~parsecs following \citep{2017patton}.  Normal neutrino mass hierarchy is assumed.}
    \label{fig:timewindow}
\end{figure}

\begin{figure}[htb!]
\gridline{\fig{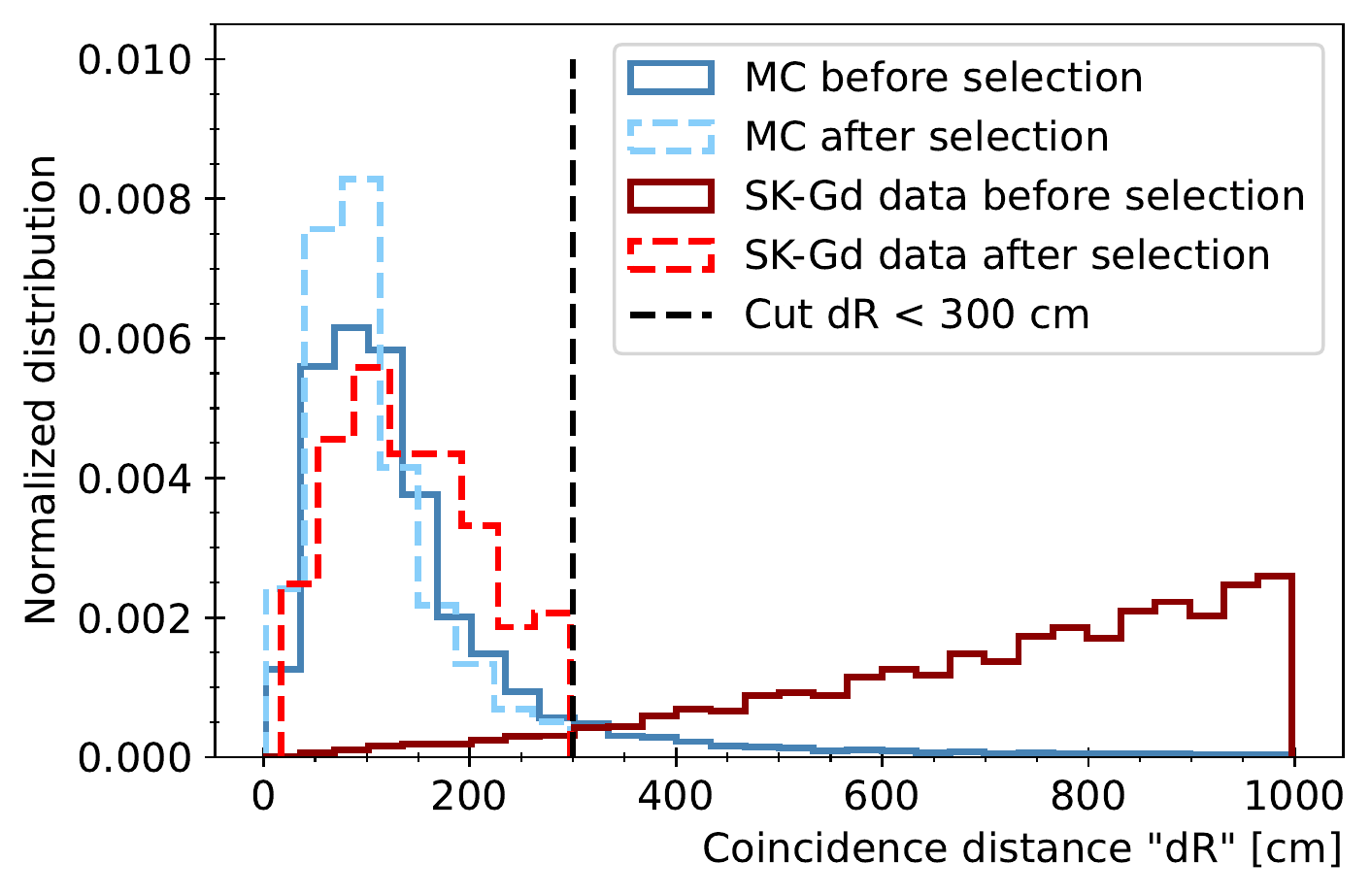}{0.45\textwidth}{(a)}
          \fig{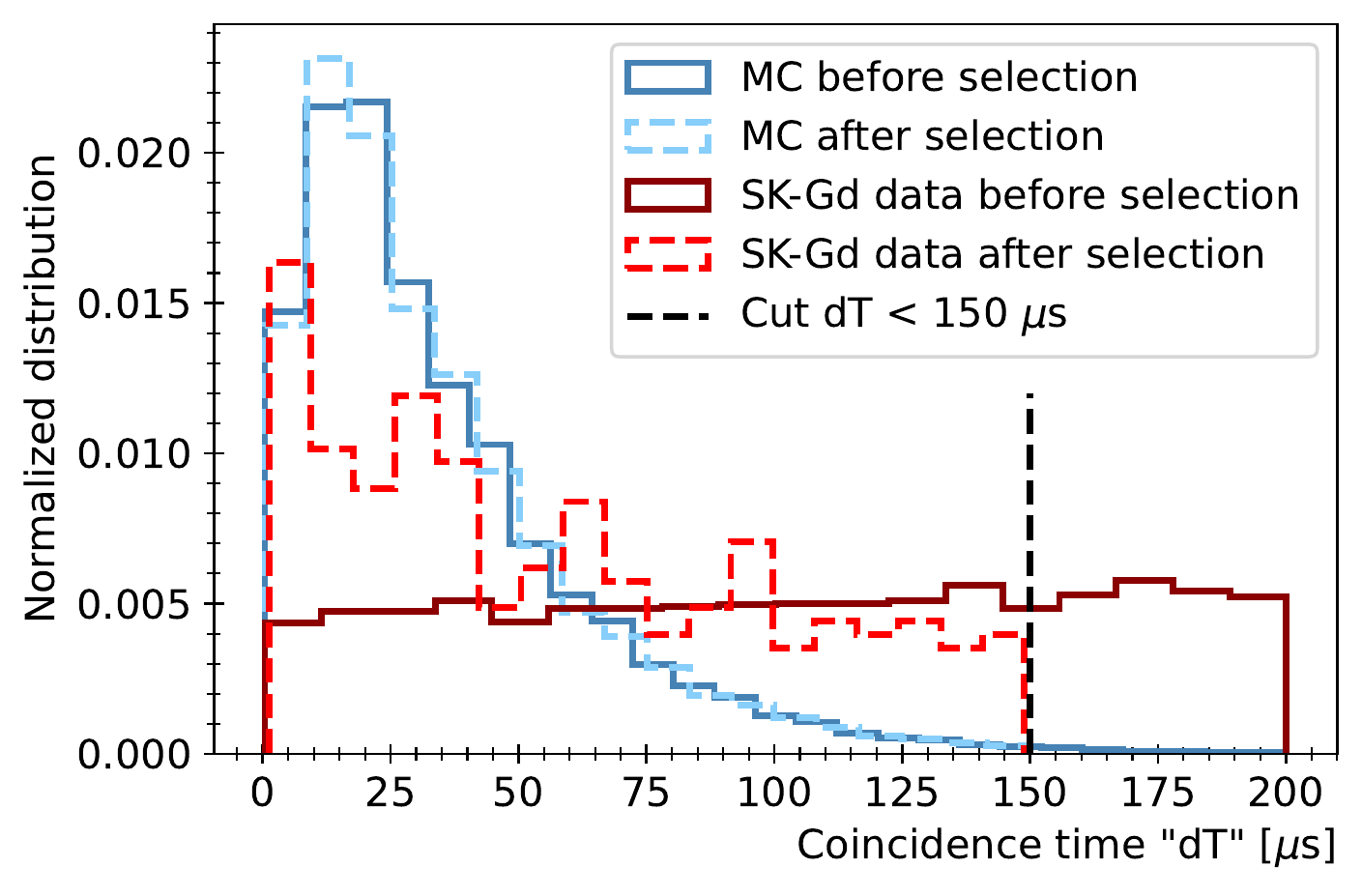}{0.45\textwidth}{(b)}
          }
\gridline{\fig{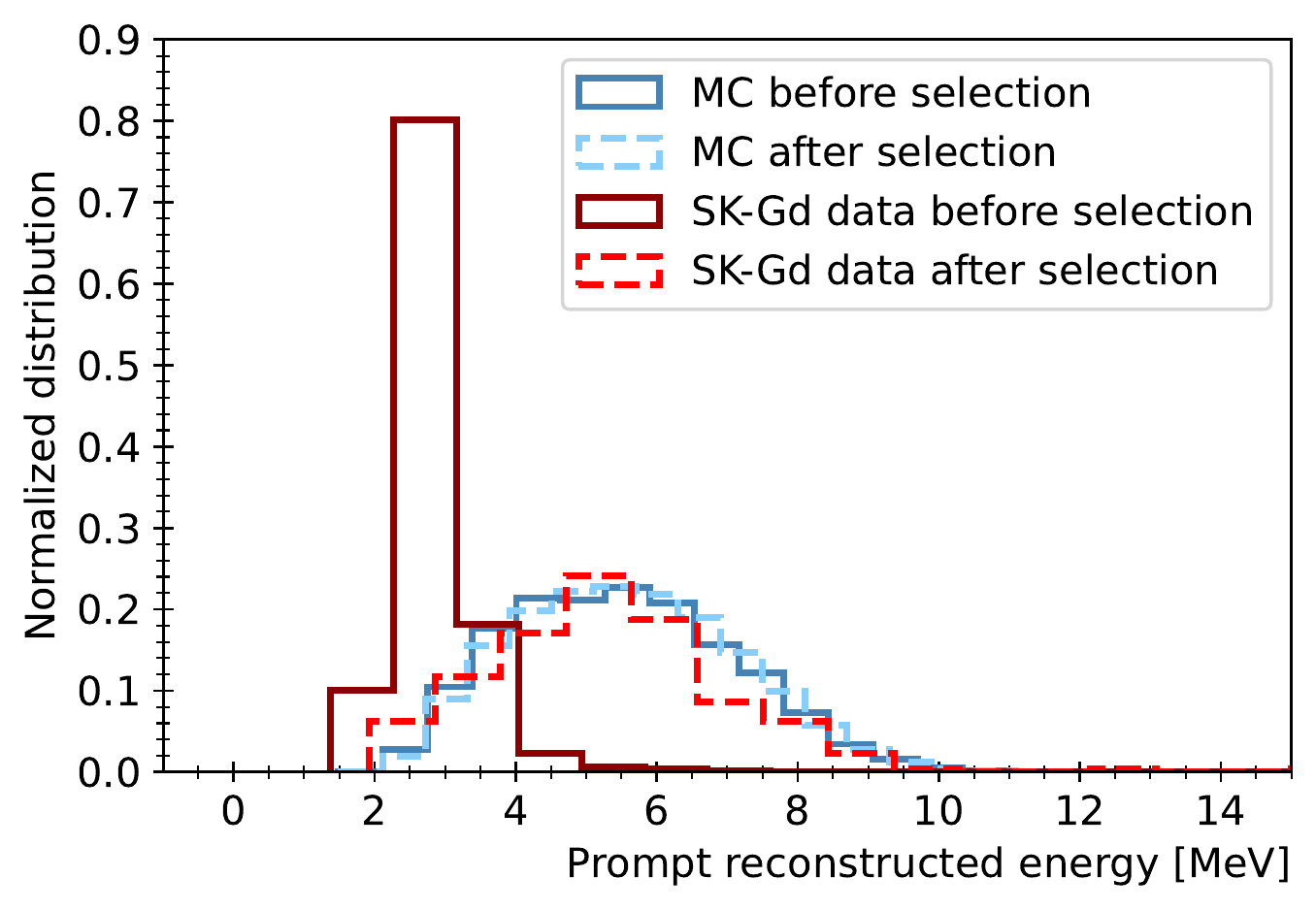}{0.45\textwidth}{(c)}
          }          
\caption{Comparison of SK-Gd data and simulation for coincidence distribution variables (a) distance dR, (b) time dT, and (c) prompt reconstructed energy. Dark colors represent the variables before the final selection and light colors after the selection. Histograms are normalised to area = 1.}
\label{fig:variables}
\end{figure}

Some important statistical parameters such as signal and background time-windows, hypothesis tests, and false positive rates need to be taken into account to estimate the sensitivities. The statistical evaluation that is performed by the alert system is also used to provide the early warnings and estimate the range of detection of pre-supernova stars. These evaluations are performed as Poissonian counting experiments~\citep{luca}. For counting experiments, the p-value is the probability of counting a total number of events greater than or equal to the expected number in which only background and no signal is present. One way to calculate the p-value of the total number of selected events inside a pre-defined time window $N_{events}$, for an expected number of background events $N_{BG}$ is:
\begin{equation}
    p = P(N_{events}  \geq N_{BG}) =  \sum_{n=N_{events}}^{\infty } Pois(n;N_{BG})
    \label{eq:pvalue}
\end{equation}
where $Pois(n;N_{BG})$ is the Poisson distribution of the variable $n$ with average $N_{BG}$.

$\alpha$-Ori (Betelgeuse in the constellation Orion) is currently the best candidate for detecting pre-supernova neutrinos. Cuts and statistical parameters were optimized to give the longest early warnings possible for Betelgeuse. Even in the best current estimations, its mass and distance from the Earth remain somewhat uncertain but are highly correlated; consequently, results are given in combinations to match what is commonly found in the literature and the available pre-supernova models: 150~pc for 15~M\textsubscript{\(\odot\)} and 220~pc for 25~M\textsubscript{\(\odot\)}~\citep{betelgeuse_ref1, betelgeuse_ref2, Joyce_reference}. The signal window chosen to perform statistical evaluations was based on which choice would maximize the warning time for Betelgeuse-like models -- the earliest alerts would be sent for a potential CCSN of Betelgeuse. Figure~\ref{fig:timewindow} shows the warning time for Betelgeuse-like models in the last 70 hours before CCSN as a  function of the signal time window used for statistical evaluations. A signal window in the interval from 7 hours to 13 hours would be one in which all Betelgeuse-like models have their longest early warnings. An window near the lower end of this range, 8~hours, but comfortably removed from the steep falloff in warning times seen with shorter windows, was chosen to maximize the early warning time while simultaneously minimizing potential negative impacts of the pre-supernova alarm inducing long interruptions in the usual data acquisition or planned calibration work in the SK detector.

\subsection{Results for SK's First Gadolinium Loading}
The following results are dependent on the neutrino mass ordering, ZAMS mass of the star, distance to the star, and star evolution model. The models for neutron capture on Gd being used are the $\gamma$-ray models GLG4sim and ANNRI, employed to give systematic uncertainties to the results. 

With the implementation of BDT$_{online}$, sensitivity estimations were improved and faster data processing was achieved, as fewer events had to be carried through the reduction after the initial search for pair candidates. Other optimization cuts to the variables used for final selection were also performed in order to maximize potential early warnings. The optimized cuts used for final selection are: coincidence distance $dR < 300$~cm, coincidence time $dT < 150$~$\mu$s, and $BDT \ score > -0.10$, resulting in an irreducible background rate of approximately $0.1$ events/hour\footnote{Background levels are periodically updated for the alert system.}. Figure~\ref{fig:variables} shows some key distributions before and after the selection for the full samples of 2000 hours of SK-Gd data and IBD generated events using the GLG4sim model. The selection with BDTs also removed any visible bursts in data, which were presumably from spallation.

Betelgeuse is the tenth brightest star in the sky. It is only about 8~million years old, but had a very quick evolution and is believed to now be close to becoming a CCSN. Table~\ref{tab:expsignal_betelgeuse} shows the expected number of IBD events for Betelgeuse in SK with 0.01\% Gd, following estimations to its mass and distance in the literature. Since the $\gamma$-ray emission models, GLG4sim and ANNRI, yield very similar results, the model GLG4sim was chosen as the baseline model for all subsequent analyses.

\begin{deluxetable}{cccc}[h!]
\tablenum{1}
\tablecaption{Number of expected IBD events for Betelgeuse in Super-Kamiokande with 0.01\% Gd in the final eight hours before becoming a core-collapse supernova. Results are shown for two different $\gamma$-ray emission models.\label{tab:expsignal_betelgeuse}}
\tablewidth{0pt}
\tablehead{\colhead{Betelgeuse-like} & \multicolumn2c{Neutrino Mass Hierarchy} & \nocolhead{}\\
\colhead{Model} & \colhead{Normal Ordering} & \colhead{Inverted Ordering}}
\startdata
\textbf{\uline{15 M\textsubscript{\(\odot\)}, 150 parsecs}} &  &  \\
Odrzywolek & 30.9 – 31.5  &  8.8 – 8.9   \\
Patton & 41.9 – 42.7 & 12.5 – 12.7 \\
\hline
\textbf{\uline{25 M\textsubscript{\(\odot\)}, 220 parsecs}} &  & \\
Odrzywolek & 25.0 – 25.5 & 7.1 – 7.2 \\
\enddata
\end{deluxetable}

\begin{figure}[htb!]
    \centering
    \includegraphics[scale=0.45]{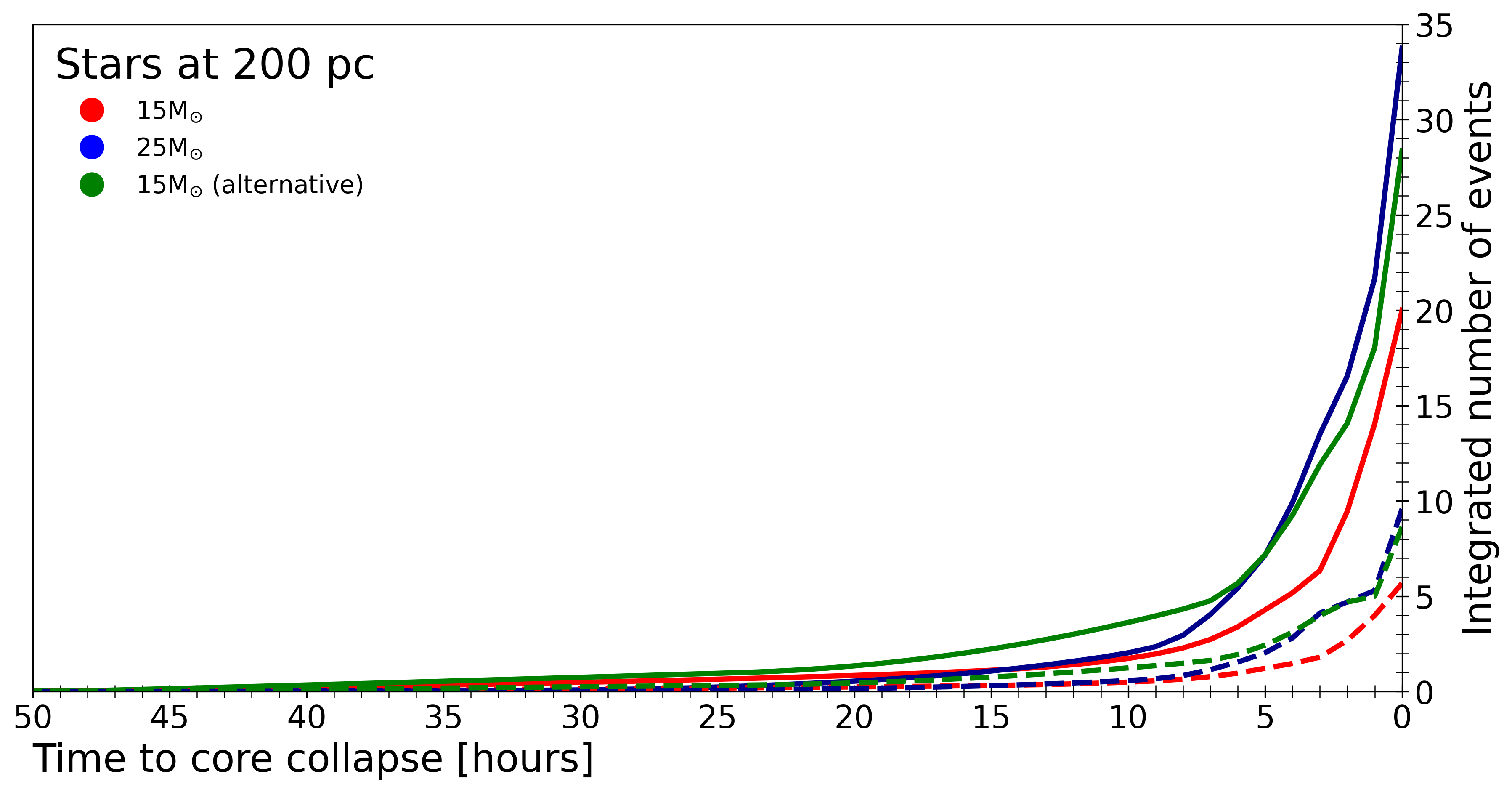}
    \caption{Integrated number of IBD events in Super-Kamiokande with 0.01\% Gd as a massive star (d = 200~pc) approaches  core-collapse. Solid lines show normal neutrino mass hierarchy and dashed lines show inverted neutrino mass hierarchy. The considered fluxes are evaluated for stars with 15~M\textsubscript{\(\odot\)} and 25~M\textsubscript{\(\odot\)} following the model \citep{Odrzywolek:2010zz}, and also alternatively for 15~M\textsubscript{\(\odot\)} stars following \citep{2017patton}.}
    \label{fig:evo_ibd_events_SK}
\end{figure}

\begin{figure}[htb!]
    \centering
    \includegraphics[scale=0.45]{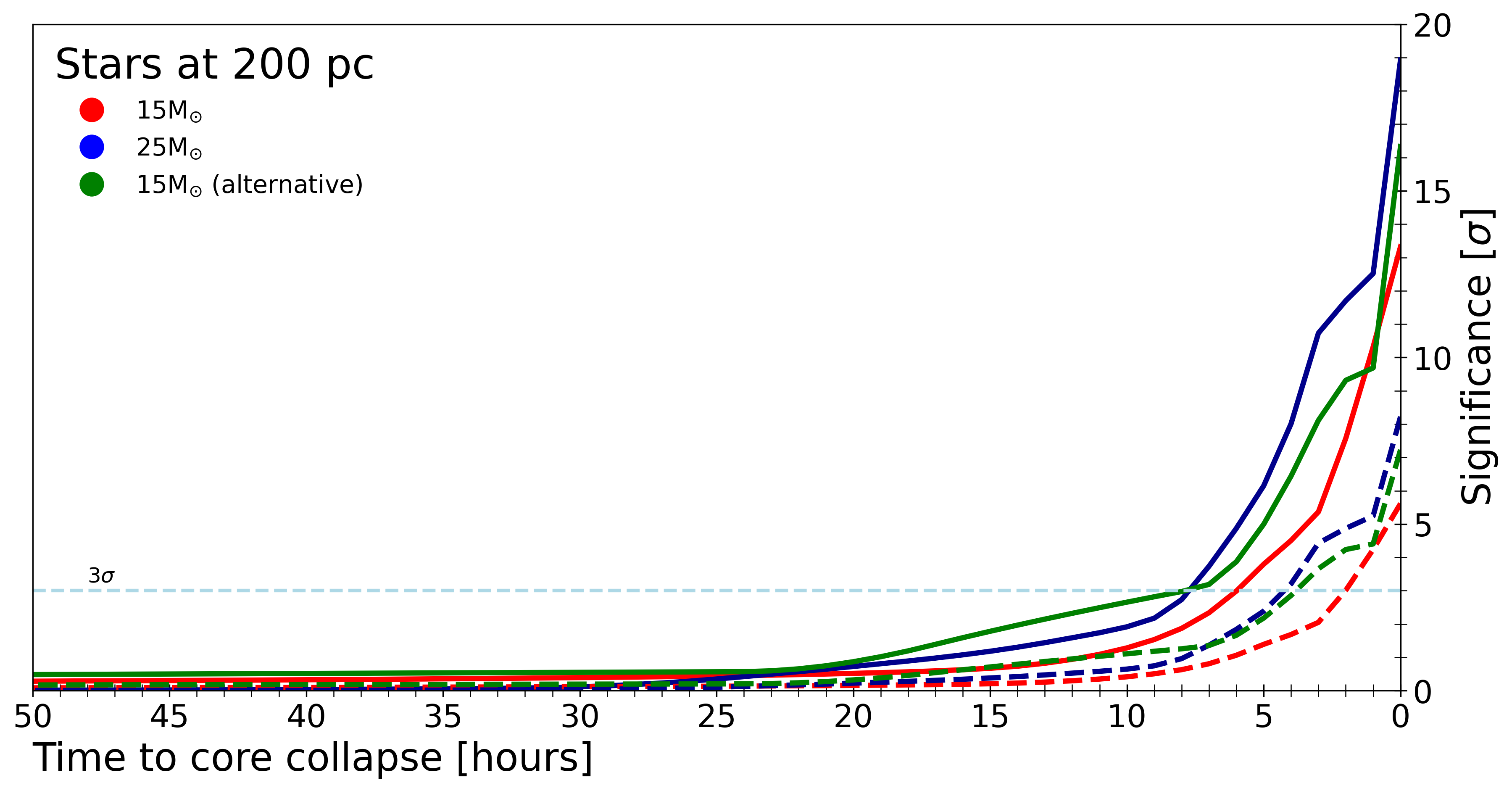}
    \caption{Evolution of the significance level in Super-Kamiokande with 0.01\% Gd over the last 50 hours before the core-collapse for massive stars at d = 200~pc. Solid lines show normal neutrino mass hierarchy and dashed lines show inverted neutrino mass hierarchy. The considered fluxes are evaluated for stars with 15~M\textsubscript{\(\odot\)} and 25~M\textsubscript{\(\odot\)} following the model \citep{Odrzywolek:2010zz}, and also alternatively for 15~M\textsubscript{\(\odot\)} stars following \citep{2017patton}.}
    \label{fig:new_evo_sig}
\end{figure}

Figure~\ref{fig:evo_ibd_events_SK} shows the evolution of the expected number of IBD events in SK-Gd with 0.01\% Gd after cuts for different pre-supernova models for stars at 200~parsecs for both normal and inverted neutrino mass hierarchies. As expected, the event rates increase significantly closer to the explosion. Figure~\ref{fig:new_evo_sig} shows the evolution of detection significance of these pre-supernova models at 200~parsecs in the hours leading up to the CCSN. 

\begin{figure}[htb!]
    \centering
    \includegraphics[scale=0.45]{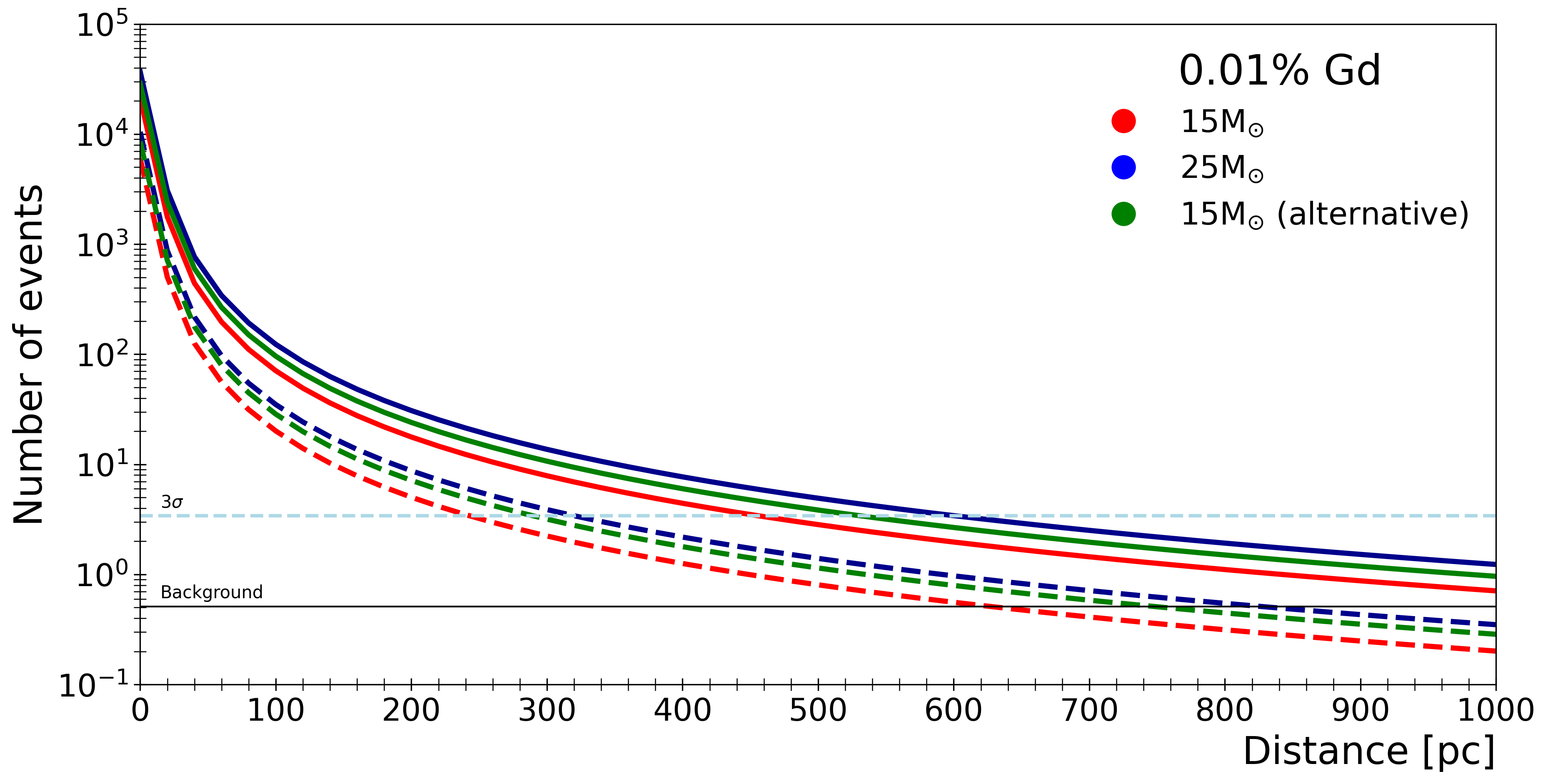}
    \caption{Expected number of IBD events in Super-Kamiokande with 0.01\% Gd as a function of distance. Solid lines show normal neutrino mass hierarchy and dashed lines show inverted neutrino mass hierarchy. The considered fluxes are integrated over eight hours preceding the core-collapse, for stars with 15~M\textsubscript{\(\odot\)} and 25~M\textsubscript{\(\odot\)} following the model \citep{Odrzywolek:2010zz}, and also alternatively for 15~M\textsubscript{\(\odot\)} stars following \citep{2017patton}.}
    \label{fig:number_ev_001}
\end{figure}

Figure~\ref{fig:number_ev_001} shows the number of IBD events expected in SK-Gd with 0.01\% Gd as a function of the distance to the star integrated over the last eight hours preceding the CCSN. Background rates are also shown. These results indicate that SK with 0.01\% Gd can observe pre-supernova neutrinos with a 3$\sigma$ detection significance for stars up to 600 parsecs away from Earth.

These results demonstrate the possibility of detecting pre-supernova neutrinos even in the first phase of SK-Gd, and also suggest the viability of creating a pre-supernova alarm based on their detection. Discussions about the development of such a pre-supernova alert system will be presented in Section~\ref{sec:alert_system}. Subsequent phases of SK-Gd will have greater Gd concentrations, and predictions for their pre-supernova sensitivities will be shown next.

\subsection{Future SK Gadolinium Loading Phases}
After characterizing backgrounds with the current SK-Gd data, evaluations of sensitivities for future Gd loading concentrations were performed. The next proposed phase is for a concentration of 0.03\% Gd, while the final phase will have a concentration of 0.10\% Gd. Since November 2020, the 200-ton EGADS detector has been loaded with 0.03\% Gd and taking data; all has gone well, verifying the feasibility of the next SK-Gd phase planned to begin in mid-2022. With higher concentrations of Gd in the SK water, neutrons will be observed with increasing efficiency, which will enhance the experiment's pre-supernova neutrino detection capabilities.

\begin{figure}[htb!]
\gridline{\fig{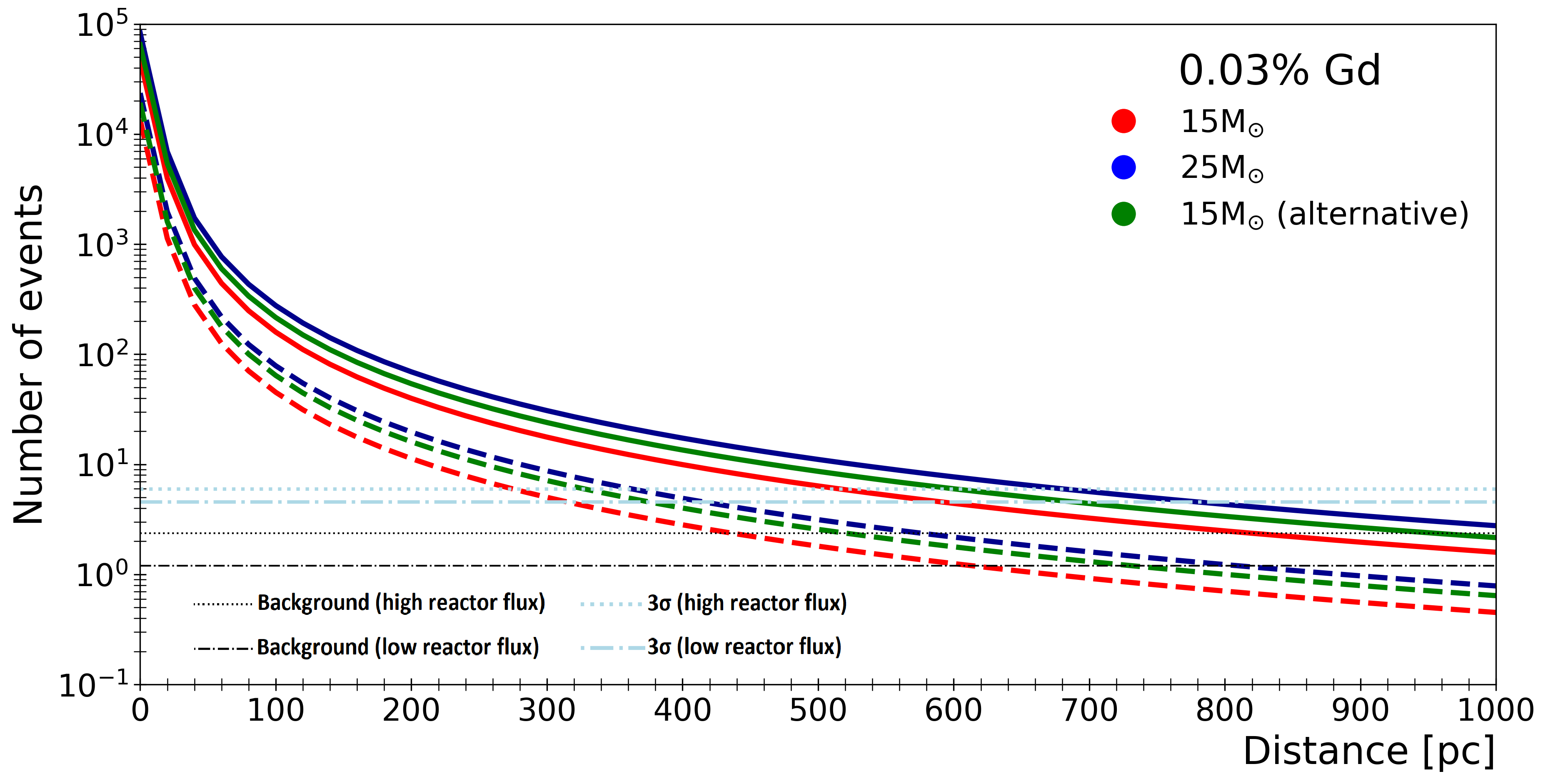}{0.5\textwidth}{(a)}
          \fig{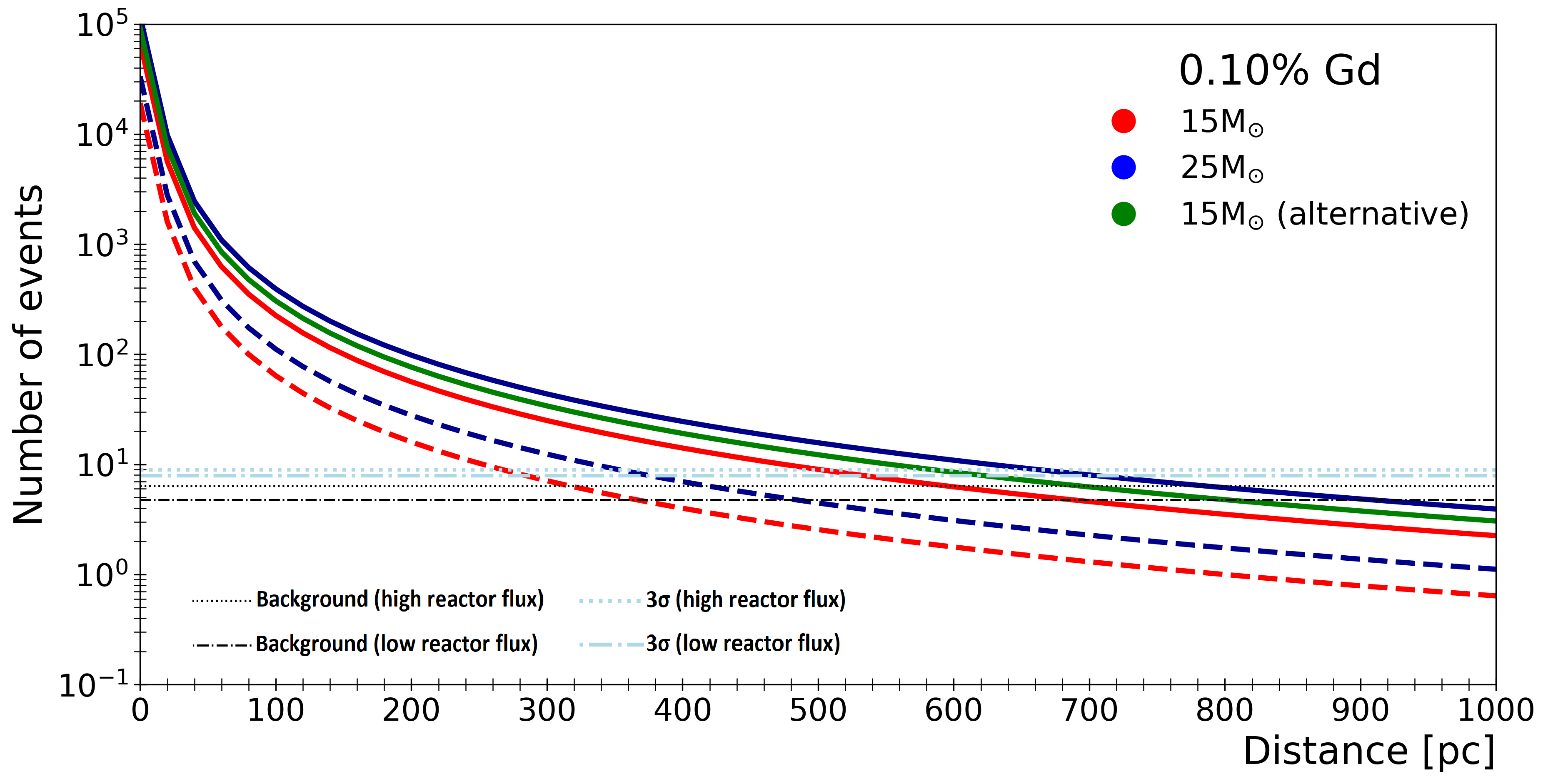}{0.5\textwidth}{(b)}
          }
\caption{Expected number of IBD events in Super-Kamiokande with (a) 0.03\% Gd and (b) 0.10\% Gd as a function of distance. Solid lines show normal neutrino mass hierarchy and dashed lines show inverted neutrino mass hierarchy. Results are shown considering two reactor scenarios: with low reactor flux (all Japanese reactors turned off), and high reactor flux (double the current contribution). The considered fluxes are integrated over eight hours preceding the core-collapse, for stars with 15~M\textsubscript{\(\odot\)} and 25~M\textsubscript{\(\odot\)} following the model \citep{Odrzywolek:2010zz}, and also alternatively for 15~M\textsubscript{\(\odot\)} stars following \citep{2017patton}.}
\label{fig:number_ev_future}
\end{figure}

Figure~\ref{fig:number_ev_future} shows the number of IBD events expected in SK integrated over the last eight hours preceding a CCSN as a function of the distance to the star for the future phases of SK-Gd. Estimated increases in the background rates are also shown. 

Backgrounds to the pre-supernova IBD signal will naturally increase as the Gd concentration is raised, since doing so will also mean higher efficiencies for detecting the IBD reactions arising from reactor and geo neutrinos, which are the main physics backgrounds to pre-supernova neutrinos.  In addition, further dissolving of Gd salts in the detector will mean greater overall radioactive contamination, increasing the rate of backgrounds from $(\alpha,n)$ and spontaneous fission processes. Due to the uncertain future of nuclear power in Japan, backgrounds are considered for two scenarios: with low reactor fluxes, in which all Japanese reactors are off, and with high reactor fluxes, meaning double the current reactor contributions. 
These preliminary results show that, in optimistic scenarios, for SK-Gd with 0.03\% the pre-supernova neutrinos can be detected from stars up to about 700~parsecs away, and for 0.10 \% Gd the detection radius  will be over 700~parsecs. As will be discussed later, future Gd loading phases should also enable earlier warnings from the alert system.

Other current and near-future neutrino experiments such as  SNO+ \citep{Albanese_2021} and JUNO \citep{juno2022} expect to have the ability to detect pre-supernova neutrino signals, while future dark matter direct detection experiments could detect all flavors of pre-supernova neutrinos via coherent neutrino-nucleus scattering \citep{Raj_2020}. The Kamioka Liquid Scintillator Antineutrino Detector (KamLAND) has an active pre-supernova alarm and comparisons with SK-Gd are shown next.

\subsection{Comparison to KamLAND}
The KamLAND experiment -- also located in the Kamioka mine -- has a working pre-supernova alert system with expected sensitivities comparable to SK-Gd. KamLAND is a liquid scintillator detector, with lower energy thresholds than SK and lower background rates. The pre-supernova alert system in KamLAND works with a set background rate of between 0.071 and 0.355 events/day, depending on the reactor activity in Japan, and integrates selected events over a 48 hour time window every 15 minutes~\citep{2016kamland}. KamLAND can be expected to see many fewer pre-supernova events than Super-Kamiokande, however, since KamLAND's fiducial volume is less than 5\% that of SK.
 
\begin{figure}[htb!]
    \centering
    \includegraphics[scale=0.45]{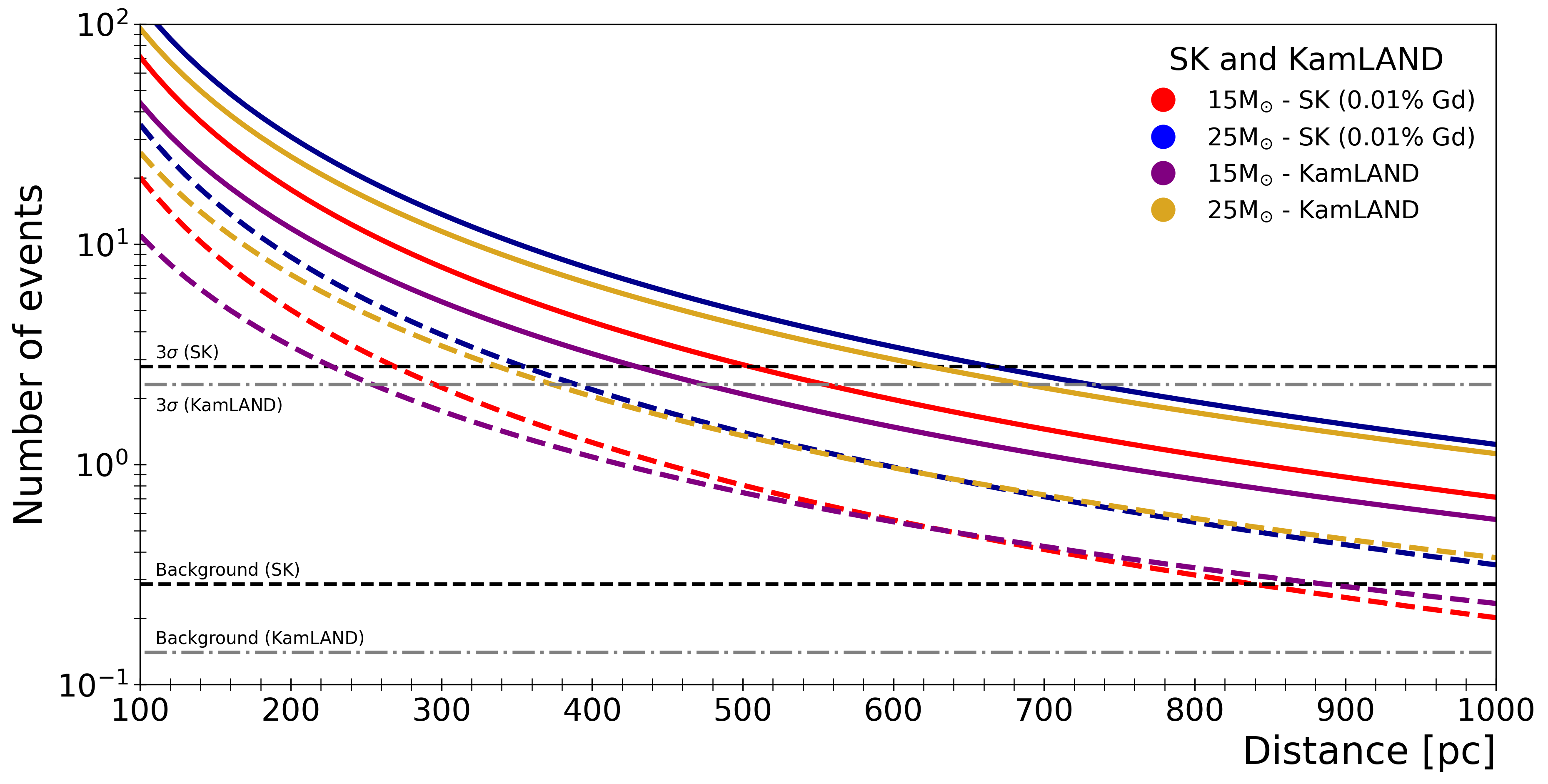}
    \caption{Expected number of IBD events in Super-Kamiokande with 0.01\% Gd and for KamLAND as a function of distance. All Japanese reactors are assumed to be off. Solid lines show normal neutrino mass hierarchy and dashed lines show inverted neutrino mass hierarchy. The considered fluxes are integrated over eight hours preceding the core-collapse for SK-Gd and 48 hours for KamLAND, for stars with 15~M\textsubscript{\(\odot\)} and 25~M\textsubscript{\(\odot\)} following the model \citep{Odrzywolek:2010zz}.}
    \label{fig:kamland_001}
\end{figure}

\begin{figure}[htb!]
\gridline{\fig{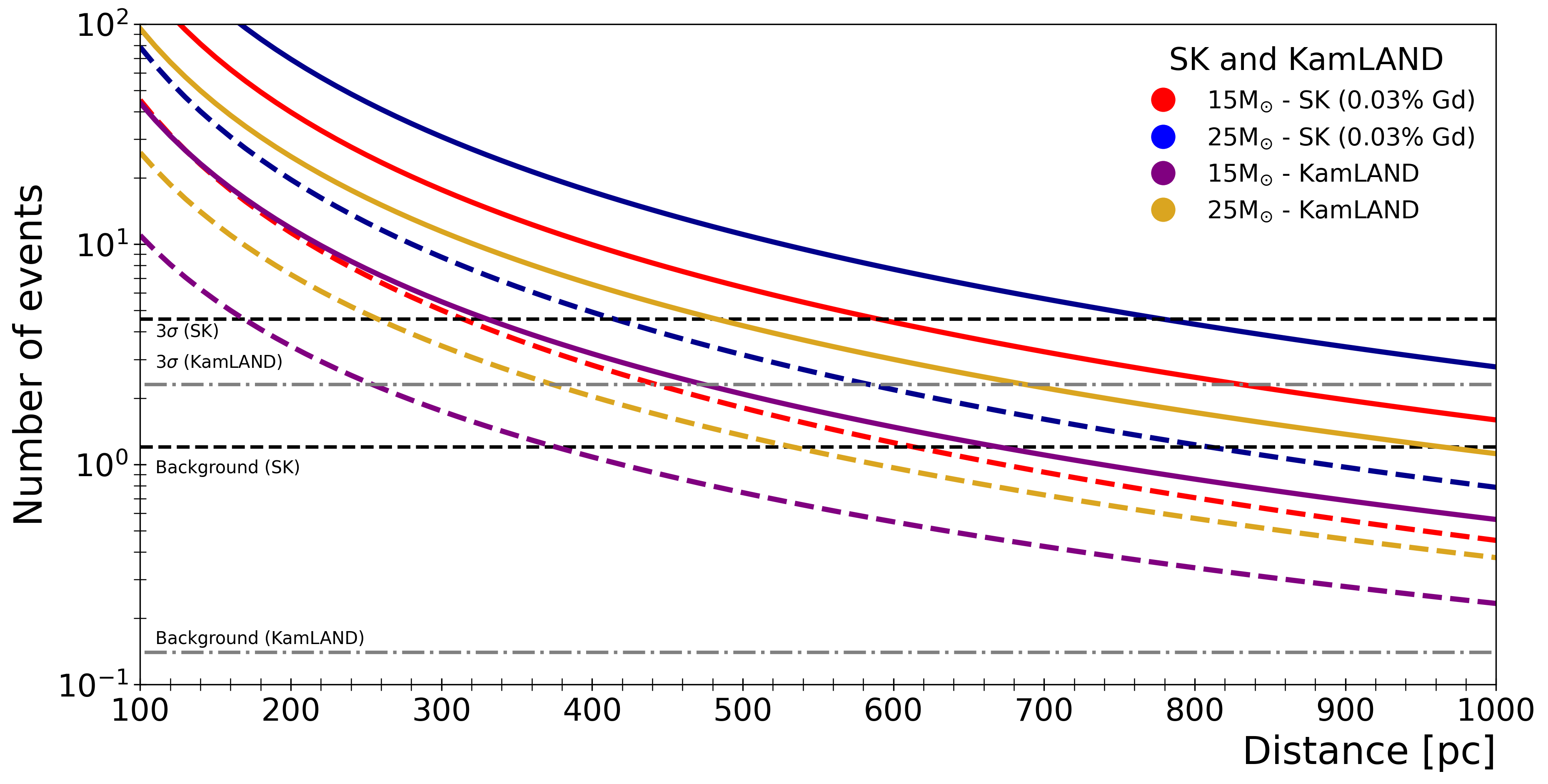}{0.49\textwidth}{(a)}
          \fig{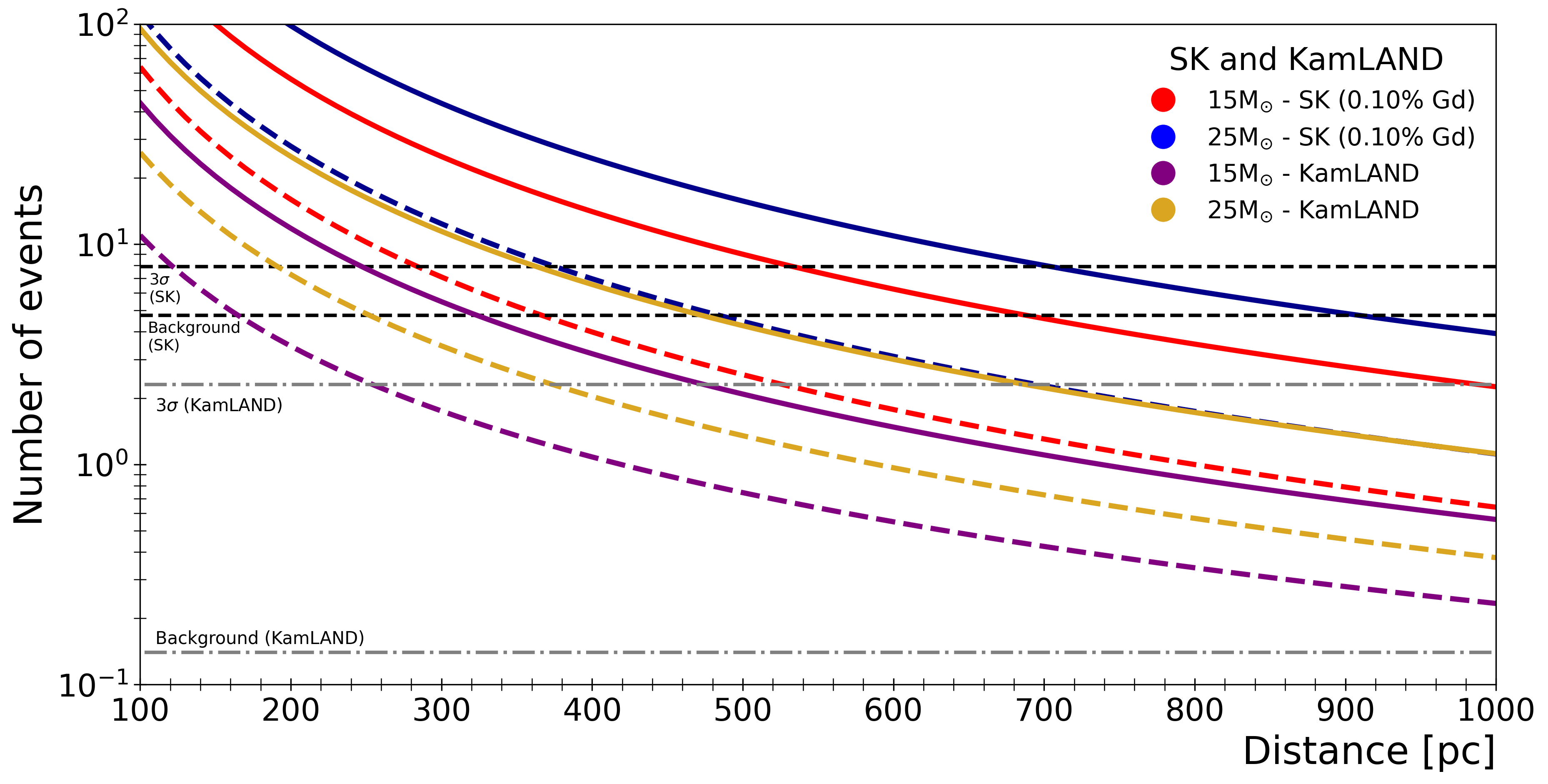}{0.49\textwidth}{(b)}
      }
\caption{Expected number of IBD events in KamLAND and in Super-Kamiokande with (a) 0.03\% Gd and (b) 0.10\% Gd as a function of distance. All Japanese reactors are assumed to be off. Solid lines show normal neutrino mass hierarchy and dashed lines show inverted neutrino mass hierarchy. The considered fluxes are integrated over eight hours preceding the core-collapse for SK-Gd and 48 hours for KamLAND, for stars with 15~M\textsubscript{\(\odot\)} and 25~M\textsubscript{\(\odot\)} following the model \citep{Odrzywolek:2010zz}.}
\label{fig:kamland_future}
\end{figure}

Based on the results reported in~\citep{2016kamland}, the relative sensitivity to pre-supernova neutrinos of the current SK-Gd phase and KamLAND is shown in Figure~\ref{fig:kamland_001}. Figure~\ref{fig:kamland_future} shows the same comparison for future SK-Gd phases.

Combining results of both detectors could lead to extended detection ranges and earlier warnings for a joint alarm; discussions between the Collaborations have been initiated with this goal in mind.
\section{Online Search for Pre-Supernova Neutrino  Signal}\label{sec:alert_system}
Super-Kamiokande's pre-supernova alert system is intended to provide early warning of an incipient nearby supernova based on the detection of pre-supernova neutrinos.  It became operational --  analyzing realtime data and ready to send out alarms of potential CCSNs -- on October 22\ts{nd}, 2021. 


The alarm software is installed in the WIT online computer cluster, performing its pre-supernova neutrino search right after the fast online vertex reconstruction, as described in Section~\ref{sec:WITtrigger}, is completed. The alarm has its own organizing processes which utilize: 1) information from the hardware trigger, a real-time counter used to group together acquired data in time, and 2) reconstructed time of events.  This allows the alarm software to calculate the absolute time of events and sort them in time.

To avoid false alerts, a volume cut is applied around calibration sources deployed inside the detector. Also, when calibration work is actively being performed in SK the system automatically ignores the data from that period by not processing any new data until the work is finished. False alerts from a statistical point of view will be discussed later in Section~\ref{sec:alarmdecision}.

\subsection{Data Flow}
The alert system works with default values of background rate and signal time window, which are parameters used for statistical evaluation. As background levels are periodically updated, it is possible to change the input values of these parameters in the system.

When active, the system checks all the available data from the WIT trigger, creating a “to-do” queue, which includes data that have not been processed yet. All data in the queue go through reduction, which starts with the search of IBD pair candidates. This is followed by a pre-selection of the pairs, which includes BDT$_{online}$. Then a final selection takes place using coincidence distance dR, coincidence time dT, and a BDT based on angular distribution of hits, reconstructed energy and quality, and distance from events to the detector wall, as explained in Section~\ref{sec:event_selection}. When the queue is empty, the system will look again for new data to create a new "to-do" queue.

All of the IBD pair candidates have their information saved: the absolute time of events, coincidence time, distance between prompt and delayed parts of the IBD event, and reconstructed vertices. This information is also used to graphically follow trends in the time evolution of candidate events and significance of detection, as well as the distribution of these events inside the SK detector. 

A rough estimation for the latency time of the alert system, which is the interval between the time when an event actually  occurs, is processed through WIT, and the alert system makes an alarm decision, is summarized in Table~\ref{tab:timing_alarm} based on tests performed on the system. Since they occur sequentially, the times in this table must be added together to arrive at the total latency, which is currently about 11 minutes. As mentioned previously, the WIT system is now being upgraded with new machines; this is expected to reduce the overall latency time to well below 10 minutes.

\begin{deluxetable}{cc}[htb!]
\tablenum{2}
\tablecaption{Estimated latency time of each step in the pre-supernova alert system. Total latency time is the sum of the the latency of each step. }\label{tab:timing_alarm}
\tablewidth{0pt}
\tablehead{\colhead{Process} & \colhead{Estimated Time}}
\startdata
Data Fitting (WIT system) & 4 minutes \\
Process Queue ($\sim 2 \times 10^{6}$ events) & 5 minutes \\
Alarm Decision/Export Results & 2 minutes \\
\enddata
\end{deluxetable}

Every time a new subset of data is processed, the system calls a function to perform the statistical evaluation, which counts the number of events that were selected inside the signal time window in order to make an alarm decision. If the decision is positive, the system sends alerts regarding the possibility of a supernova. 

The Super-Kamiokande detector is monitored constantly by collaborators. The status of the pre-supernova alert system is included in the SK monitoring via a web page which displays  realtime information from the system. In addition to technical status-related data, the web page includes the latest results from (and history of) the alert system's statistical evaluations. These  graphs of the recent IBD pair candidate rate and evolution of the resulting significance level can be used to monitor unusual fluctuations, in particular sudden increments in IBD pair candidates and detection significance which can lead to potential alerts.

\subsection{Alarm Decision}\label{sec:alarmdecision}
The alarm decision determines whether the software should send an alert or not. Many factors have to be taken into account, such as the current status of the detector, background rate levels, and false positive rates. 

After processing new data, the running number of selected IBD events undergoes a statistical evaluation to determine the significance of the detection, calculated with Poisson statistics as discussed in Section~\ref{sec:sensitivity}. When testing a hypothesis, one possible outcome is a false positive, which is the probability of believing a condition exists when it does not.

The choice of an acceptable false positive rate (FPR) is often related to an experiment's sensitivity, since requiring greater confidence in a result in order to reduce the FPR means less chance of positively identifying more marginal, but nevertheless real, signals. Consequently, a decrease in the false positive rate implies an increase in the experiment's false negative rate (FNR), and vice versa.

In order to have a real-world basis for the pre-supernova alert system's chosen FPR, it was decided that the rate of core-collapse supernova explosions in our galaxy should serve as the maximum FPR.  In other words, statistically speaking the system should not issue false alerts more frequently than Milky Way (MW) core-collapse explosions are expected to take place.  Of course, the true rate of detectable pre-supernova explosions will be considerably lower than this since they must be located much closer to Earth than the average MW explosion, but the trade-off between FPR and FNR means that allowing the occasional false positive will improve sensitivity to the true positives.

From \citep{Adams:2013ana} the rate of galactic core-collapse supernova explosions is taken to be $3.2_{-2.6}^{+7.3}$ per century.  This rate is derived using all supernova events observed over the last millennium with various corrections applied.  The most important correction is due to what fraction of the galaxy is visible to the naked eye, since all recorded explosions inside the MW were observed before the invention of the telescope. 

To introduce this factor, inside the signal window of eight hours chosen to perform statistical evaluations a cut-off is applied to the p-value: $\frac{3.2}{century}$ per 8 hours $\sim 7.4 \times 10^{-5}$, requiring the alarm to send alerts in case the number of IBD events inside the signal window are enough to give significance levels above $\sim 4\sigma$. However, if the significance level exceeds 3$\sigma$ then a preliminary alert is sent to SK experts so they may begin monitoring the situation as
it develops.

\begin{figure}[htb!]
    \centering
    \includegraphics[scale=0.65]{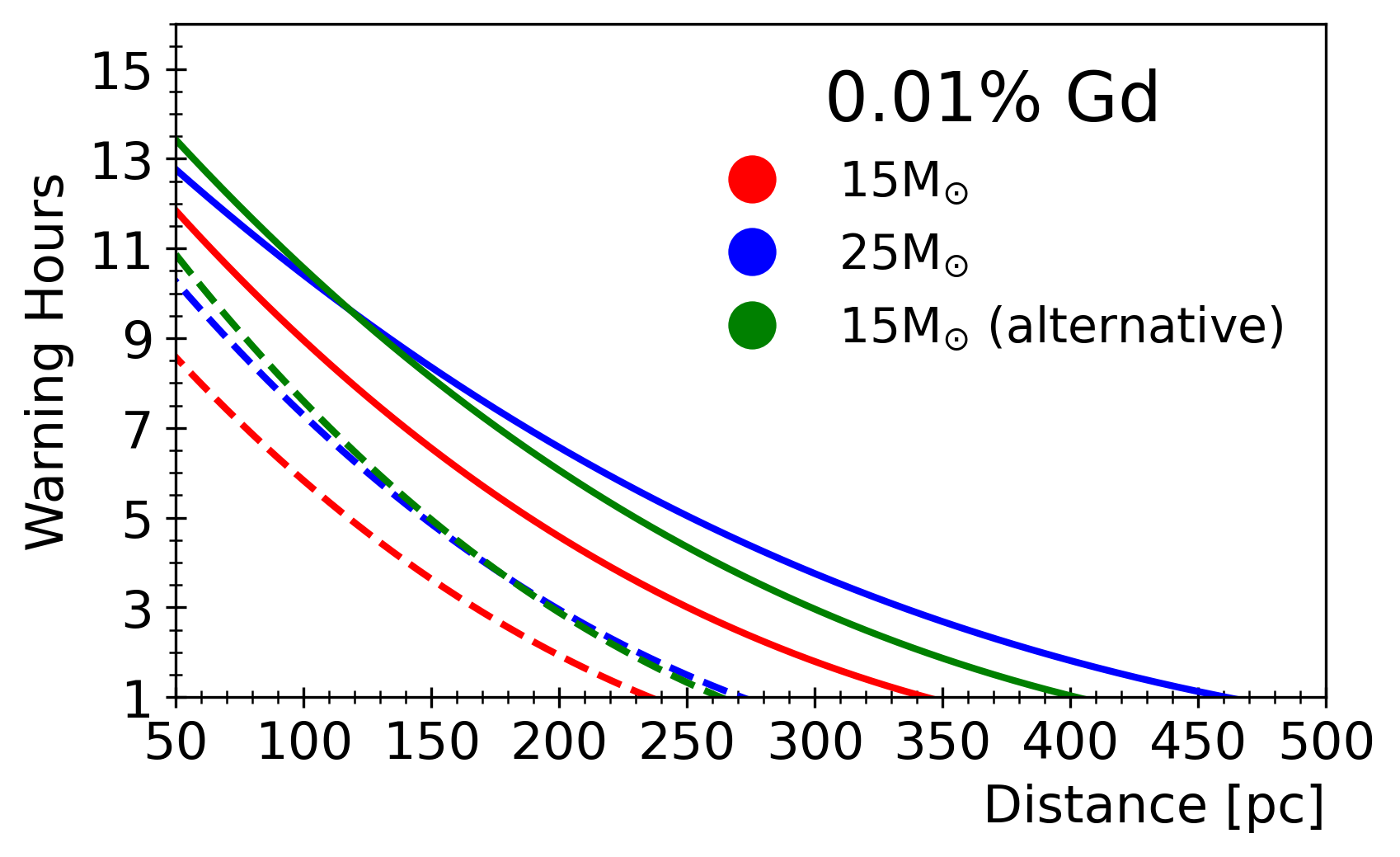}
    \caption{Expected warning time as function of distance for a 3$\sigma$ detection by the pre-supernova alert system for Super-Kamiokande with 0.01\% Gd. Solid lines show normal neutrino mass hierarchy and dashed lines show inverted neutrino mass hierarchy. The considered fluxes are evaluated for stars with 15~M\textsubscript{\(\odot\)} and 25~M\textsubscript{\(\odot\)} following the model \citep{Odrzywolek:2010zz}, and also alternatively for 15~M\textsubscript{\(\odot\)} stars following \citep{2017patton}.}
    \label{fig:warning_001}
\end{figure}

\begin{deluxetable}{ccccccc}
\tablenum{3}
\tablecaption{List of likely pre-supernova candidates with estimated masses and distances.}
\tablewidth{800pt}
\tabletypesize{\scriptsize}
\tablehead{
\colhead{Catalog Name} & \colhead{Common Name} & \colhead{Constellation} & \colhead{Distance} & \colhead{Mass}  \\ 
\colhead{}				& \colhead{} &				\colhead{} &			\colhead{(kpc)}  & \colhead{(M\textsubscript{\(\odot\)})}
} 
\startdata
HD 116658 &  Spica/\(\alpha\)  Virginis &  Virgo &  $0.0766 \pm 0.0041 {}^{(a)}$ &  ${11.43}^{+1.15}_{-1.15} {}^{(b)}$ \\
HD 149757 &  \(\zeta\)  Ophiuchi &  Ophiuchus &  $0.107 \pm 0.004 {}^{(c)}$ &  ${13}^{+10}_{-6} {}^{(d)}$ \\
HD 129056 &  \(\alpha\)  Lupi &  Lupus &  $0.143 \pm 0.003 {}^{(a)}$ &  ${10.1}^{+1.0}_{-1.0} {}^{(a)}$ \\
HD 78647 &  \(\lambda\)  Velorum &  Vela &  $0.167 \pm 0.003 {}^{(a)}$ &  ${7.0}^{+1.5}_{-1.0} {}^{(a)}$ \\
HD 148478 &  Antares/\(\alpha\)  Scorpii &  Scorpius &  $0.169 \pm 0.030 {}^{(a)}$ &  ${11.0-14.3} {}^{(a)}$ \\
HD 39801 &  Betelgeuse/\(\alpha\)  Orionis &  Orion &  $0.168^{+0.027}_{-0.015} {}^{(e)}$ &  ${16.5}-{19} {}^{(e)}$ \\
HD 206778 &  \(\epsilon\)  Pegasi &  Pegasus &  $0.211 \pm 0.006 {}^{(a)}$ &  ${11.7}^{+0.8}_{-0.8} {}^{(a)}$ \\
HD 89388 &  q Car/V337 Car &  Carina &  $0.230 \pm 0.020 {}^{(a)}$ &  ${9.0} \pm 1.6 {}^{(f)}$ \\
HD 34085 &  Rigel/\(\beta\)  Orion &  Orion &  $0.237^{+0.057}_{-0.038} {}^{(g)}$ &  ${21.0}^{+3.0}_{-3.0} {}^{(a)}$ \\
HD 210745 &  \(\zeta\)  Cephei &  Cepheus &  $0.256 \pm 0.006 {}^{(a)}$ &  ${10.1}^{+0.1}_{-0.1} {}^{(a)}$ \\
HD 200905 &  $xi$ Cygni &  Cygnus &  $0.278 \pm 0.029 {}^{(a)}$ &  $8.0 {}^{(a)}$ \\
HD 47839 &  S Monocerotis A &  Monoceros &  $ 0.282 \pm 0.040 {}^{(a)}$ &  $29.1 {}^{(a)}$ \\
HD 47839 &  S Monocerotis B &  Monoceros &  $ 0.282 \pm 0.040 {}^{(a)}$ &  $21.3 {}^{(a)}$ \\
HD 93070 &  w Car/V520 Car &  Carina &  $0.294 \pm 0.023 {}^{(a)}$ &  ${7.9}^{+0.1}_{-0.1} {}^{(a)}$\\
HD 68553 &  NS Puppis &  Puppis &  $0.321 \pm 0.032 {}^{(a)}$ &  $9.7 {}^{(a)}$ \\
HD 36389 &  CE Tauri/119 Tauri &  Taurus &  $0.326 \pm 0.070 {}^{(a)}$ &  ${14.37}^{+2.00}_{-2.77} {}^{(a)}$ \\
HD 68273 &  \(\gamma\)${}^2$  Velorum &  Vela &  $0.379 \pm 0.004 {}^{(h)}$ &  ${9.0}^{+0.6}_{-0.6} {}^{(a)}$ \\
HD 50877 &  \o${}^1$ Canis Majoris &  Canis Major &  $0.394 \pm 0.052 {}^{(a)}$ &  ${7.83}^{+2.0}_{-2.0} {}^{(a)}$ \\
HD 52877 &  \(\sigma\)  Canis Majoris &  Canis Major &  $0.513 \pm 0.108 {}^{(a)}$ &  ${12.3}^{+0.1}_{-0.1} {}^{(a)}$ \\
HD 208816 &  VV Cephei &  Cepheus &  $0.599 \pm 0.083 {}^{(a)}$ &  ${10.6}^{+1.0}_{-1.0} {}^{(a)}$ \\
HD 203338 &  V381 Cephei &  Cepheus &  $0.631 \pm 0.086 {}^{(a)}$ &  $12.0 {}^{(a)}$ \\
HD 17958 &  HR 861 &  Cassiopeia &  $0.639 \pm 0.039 {}^{(a)}$ &  ${9.2}^{+0.5}_{-0.5} {}^{(a)}$  \\
HD 80108 &  HR 3692 &  Vela &  $0.650 \pm 0.061 {}^{(a)}$ &  ${12.1}^{+0.2}_{-0.2} {}^{(a)}$ \\
\enddata
\tablecomments{ \\
a: \citep{articleMukhopadhyay} (and references therein) \\
b: \citep{Nieva} \\
c: \citep{Renzo} \\
d: \citep{Marcolino} \\
e: \citep{Joyce_reference} \\
f: \citep{Kallinger} \\
g: \citep{Chesneau} \\
h: \citep{10.1093/mnras/staa1290}
}
\label{tab:expsignal}
\end{deluxetable}

\subsection{Alert System Sensitivity}
The expected warning time as function of distance for a 3$\sigma$ detection by SK-Gd with 0.01\% Gd is shown in Figure~\ref{fig:warning_001}. In the most optimistic scenario, in which Betelgeuse's true mass is 15~M\textsubscript{\(\odot\)} at a distance of 150 parsecs and with normal neutrino mass hierarchy, following the model from \citep{2017patton} the alert would be about 9 hours before the core-collapse supernova. Table~\ref{tab:expsignal} lists a number of nearby stars at different distances which could potentially provide observable pre-supernova neutrino signals. 
\subsubsection{Sensitivity for Future SK-Gd Phases}
\begin{figure}[htb!]
\gridline{\fig{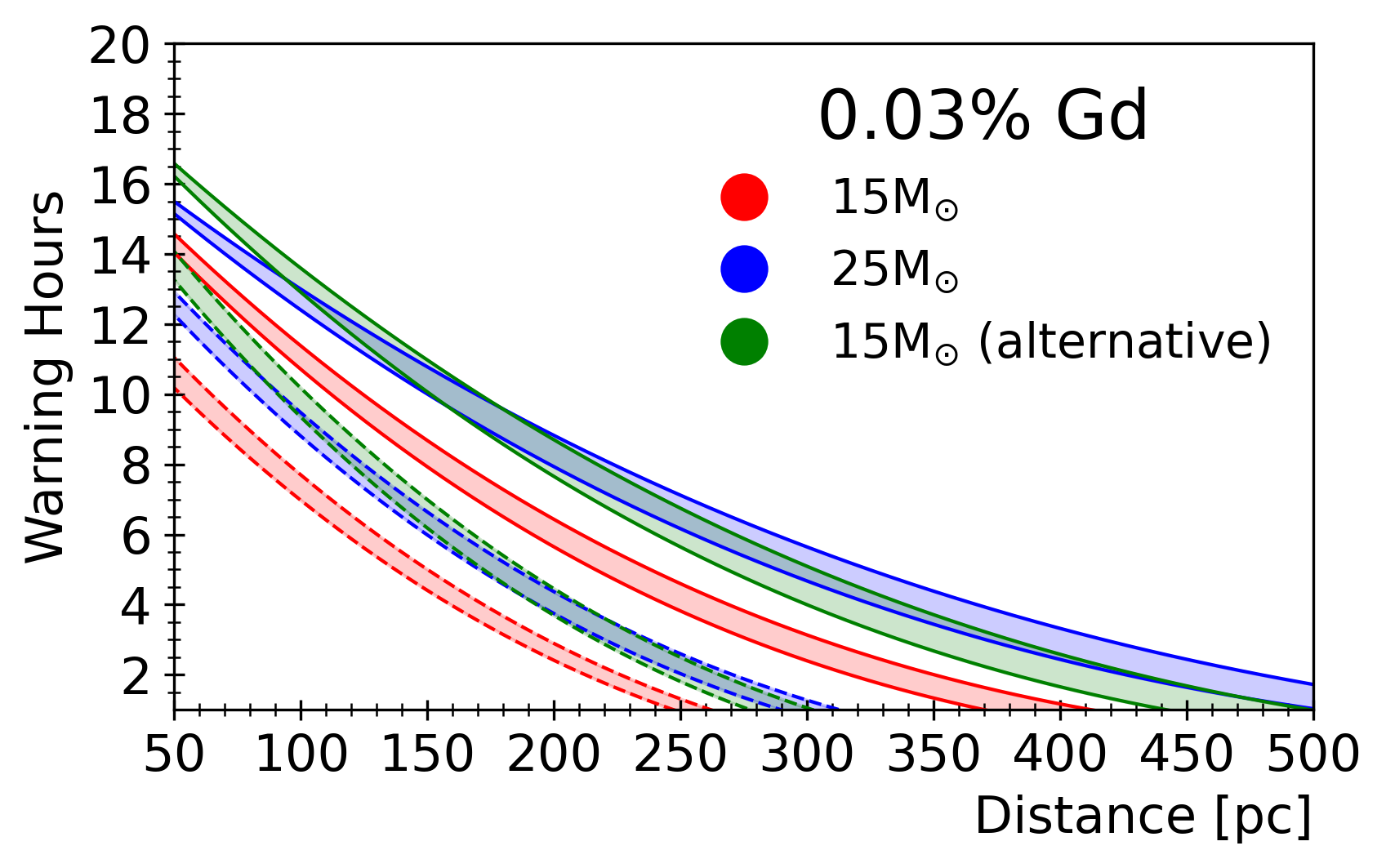}{0.49\textwidth}{(a)}
          \fig{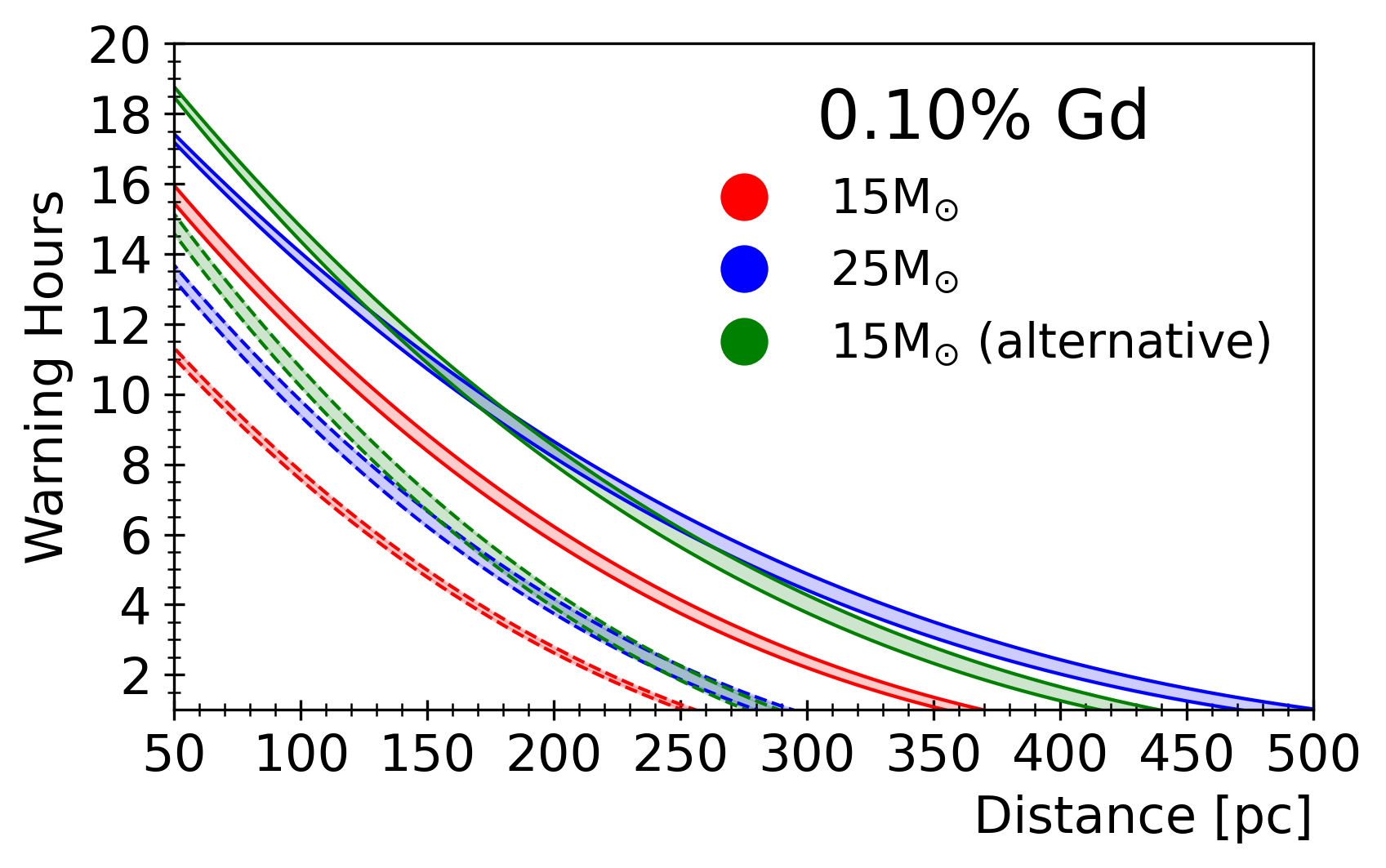}{0.49\textwidth}{(b)}
          }
\caption{Expected warning time as function of distance for a 3$\sigma$ detection by the pre-supernova alert system for Super-Kamiokande with (a) 0.03\% Gd and (b) 0.10\% Gd. The edges of the bands show two background scenarios considering low reactor flux (All Japanese reactors off) and high reactor flux (double the current contribution). Solid lines show normal neutrino mass hierarchy and dashed lines show inverted neutrino mass hierarchy. The considered fluxes are evaluated for stars with 15~M\textsubscript{\(\odot\)} and 25~M\textsubscript{\(\odot\)} following the model \citep{Odrzywolek:2010zz}, and also alternatively for 15~M\textsubscript{\(\odot\)} stars following \citep{2017patton}.}
\label{fig:warning_future}
\end{figure}

The expected early warning times for future SK-Gd phases were also evaluated after characterizing backgrounds using SK-Gd data. Although -- as previously discussed -- the background rates will be higher during the future phases, the expected sensitivity will also increase. Figure~\ref{fig:warning_future} shows the expected warning times at which the system would send alerts as function of distance for a 3$\sigma$ detection for SK with 0.03\% and 0.10\% Gd. 

During future phases of SK-Gd and for the same optimistic scenario discussed above, Betelgeuse would have early warning alerts about 10 to 12 hours in advance of its end as a core-collapse supernova. 

\section{Conclusion}
Super-Kamiokande is a neutrino observatory in operation since 1996 that has recently entered its SK-Gd phase; gadolinium was loaded into the detector in 2020, achieving a concentration of 0.01\% Gd. This phase is characterized by increased sensitivity to thermal neutrons due to the enormous capture cross section of gadolinium,
which then emits easily detectable $\gamma$-ray cascades of about 8 MeV total energy. In turn, the highly visible neutron captures made possible by Gd loading greatly enhance the detector's sensitivity to low energy electron anti-neutrinos $\bar{\nu}_e$, as their dominant interaction channel is via inverse beta decay which yields a positron and neutron in the final state. Consequently, the  Gd-loaded Super-Kamiokande now has the potential to detect as-yet-unobserved neutrinos from different astronomical sources such as the diffuse supernova neutrino background and pre-supernova stars.

Stars with greater than 8 solar masses and a growing iron core are in the final stages of fusion shortly before core-collapse; these are commonly known as pre-supernova stars. Their main cooling mechanism is the production of low energy neutrinos and anti-neutrinos from thermal and weak nuclear processes. Some of these neutrinos have enough energy to exceed the inverse beta decay energy threshold, and -- if the star is close enough -- can therefore be detected at SK-Gd. Pre-supernova models from \citep{Odrzywolek:2010zz} and \citep{2017patton} were used to evaluate the sensitivity of Super-Kamiokande with 0.01\%  gadolinium to pre-supernova neutrinos. For this gadolinium concentration, estimations showed that pre-supernova stars can be observed in SK-Gd up to an optimistic distance of 600 parsecs away from Earth.

The emission of neutrinos from pre-supernova stars occurs for hours before the core collapses, which could provide an early warning for potential supernova events. A pre-supernova alert system was developed for Super-Kamiokande and it has been active since October 22\ts{nd}, 2021. In the case of $\alpha$-Ori (Betelgeuse) with optimistic parameters, Super-Kamiokande would produce alerts up to 9 hours before the core-collapse supernova. Nearly 20 other nearby stars could have an early warning of their coming explosion via the detection of their pre-supernova neutrinos. Estimations showed that for future phases of Super-Kamiokande with increased concentrations of gadolinium, detection ranges and early alert times of the pre-supernova alarm would be extended.

\begin{acknowledgments}
\section*{Acknowledgements}
We gratefully acknowledge the cooperation of the Kamioka Mining and Smelting Company.
The Super‐Kamiokande experiment has been built and operated from funding by the 
Japanese Ministry of Education, Culture, Sports, Science and Technology; the U.S.
Department of Energy; and the U.S. National Science Foundation. Some of us have been 
supported by funds from the National Research Foundation of Korea NRF‐2009‐0083526
(KNRC) funded by the Ministry of Science, ICT, and Future Planning and the Ministry of
Education (2018R1D1A1B07049158, 2021R1I1A1A01059559); 
the Japan Society for the Promotion of Science; the National
Natural Science Foundation of China under Grants No.11620101004; the Spanish Ministry of Science, 
Universities and Innovation (grant PGC2018-099388-B-I00); the Natural Sciences and 
Engineering Research Council (NSERC) of Canada; the Scinet and Westgrid consortia of 
Compute Canada; the National Science Centre (UMO-2018/30/E/ST2/00441) and the Ministry 
of Education and Science (DIR/WK/2017/05), Poland; 
the Science and Technology Facilities Council (STFC) and GridPPP, UK; the European Union's 
Horizon 2020 Research and Innovation Programme under the Marie Sklodowska-Curie grant
agreement no.754496, H2020-MSCA-RISE-2018 JENNIFER2 grant agreement no.822070, and 
H2020-MSCA-RISE-2019 SK2HK grant agreement no. 872549.
For the purposes of open access, the authors have applied a creative
commons attribution (CC BY) licence to any author accepted manuscript
version arising.
\end{acknowledgments}

\bibliography{main.bib}{}
\bibliographystyle{aasjournal}

\end{document}